\documentclass[reqno,11pt]{amsart}
\usepackage[utf8]{inputenc}
\usepackage{enumitem}
\usepackage{esint}
\usepackage{graphicx}
\usepackage{amscd}
\usepackage{slashed}
\usepackage{amssymb}
\usepackage{mathtools} 
\usepackage[usenames,dvipsnames]{pstricks}
\usepackage[mathscr]{eucal}
\numberwithin{equation}{section}
\allowdisplaybreaks[1]

\title[Fermionic Fock Spaces and Quantum States for Causal Fermion Systems]
{Fermionic Fock Spaces and Quantum States \\
for Causal Fermion Systems}

\author[F.\ Finster]{Felix Finster}
\address{Fakult\"at f\"ur Mathematik \\ Universit\"at Regensburg \\ D-93040 Regensburg \\ Germany}
\email{finster@ur.de}

\author[N.\ Kamran]{Niky Kamran \\ \\ January / August 2021}
\address{Department of Mathematics and Statistics \\ McGill University \\ Montr{\'e}al \\ Canada}
\email{nkamran@math.mcgill.ca}

\newtheorem{Def}{Definition}[section]
\newtheorem{Thm}[Def]{Theorem}
\newtheorem{Prp}[Def]{Proposition}

\newcommand{\Thanks}{\vspace*{.5em} \noindent \thanks}
\newcommand{\beq}{\begin{equation}}
\newcommand{\eeq}{\end{equation}}
\newcommand{\Proof}{\begin{proof}}
\newcommand{\QED}{\end{proof} \noindent}

\newcommand{\la}{\langle}
\newcommand{\ra}{\rangle}

\newcommand{\bra}{\mathopen{<}}
\newcommand{\ket}{\mathclose{>}}
\newcommand{\Sl}{\mathopen{\prec}}
\newcommand{\Sr}{\mathclose{\succ}}

\newcommand{\C}{\mathbb{C}}
\newcommand{\R}{\mathbb{R}}
\newcommand{\1}{\mbox{\rm 1 \hspace{-1.05 em} 1}}

\newcommand{\N}{\mathbb{N}}

\renewcommand{\H}{\mathscr{H}}
\newcommand{\h}{\mathfrak{h}}
\newcommand{\U}{{\rm{U}}}

\newcommand{\G}{\mathscr{G}}

\newcommand{\bep}{\begin{pmatrix}}
\newcommand{\enp}{\end{pmatrix}}

\renewcommand{\O}{\mathscr{O}}

\newcommand{\F}{{\mathscr{F}}}
\newcommand{\reg}{{\text{\rm{reg}}}}

\renewcommand{\O}{{\mathscr{O}}}
\renewcommand{\L}{{\mathcal{L}}}
\newcommand{\Sact}{{\mathcal{S}}}
\newcommand{\s}{{\mathfrak{s}}}

\newcommand{\Lin}{\text{\rm{L}}}
\newcommand{\T}{{\mathscr{T}}}

\newcommand{\Fock}{{\mathcal{F}}}
\newcommand{\fermi}{{\mathrm{f}}}
\newcommand{\bose}{{\mathrm{b}}}

\newcommand{\scrM}{\myscr M}

\newcommand{\scrW}{\mathscr{W}}
\newcommand{\A}{\myscr A}

\newcommand{\J}{\mathfrak{J}}
\newcommand{\Jin}{\mathfrak{J}^\text{\rm{\tiny{in}}}}
\newcommand{\Jlin}{\mathfrak{J}^\text{\rm{\tiny{lin}}}}
\newcommand{\Jtest}{\mathfrak{J}^\text{\rm{\tiny{test}}}}
\newcommand{\Jvary}{\mathfrak{J}^\text{\rm{\tiny{vary}}}}

\newcommand{\Jfermi}{\mathfrak{J}^\text{\rm{\tiny{f}}}}

\newcommand{\Gtest}{\Gamma^\text{\rm{\tiny{test}}}}

\newcommand{\Glin}{\Gamma^\text{\rm{\tiny{lin}}}}
\newcommand{\Gdiff}{\Gamma^\text{\rm{\tiny{diff}}}}
\newcommand{\Gfermi}{\Gamma^\text{\rm{\tiny{f}}}}

\newcommand{\dyn}{{\text{\rm{dyn}}}}
\newcommand{\Ctest}{C^\text{\rm{\tiny{test}}}}
\newcommand{\Jdiff}{\mathfrak{J}^\text{\rm{\tiny{diff}}}}

\newcommand{\scrU}{{\mathscr{U}}}

\newcommand{\scrA}{{\mathscr{A}}}

\renewcommand{\div}{{\rm{div}}\,}

\newcommand{\x}{\mathbf{x}}

\newcommand{\hol}{\text{\rm{hol}}}
\newcommand{\ah}{\text{\rm{ah}}}
\newcommand{\symm}{\text{s}}

\newcommand{\itemD}{\item[{\raisebox{0.125em}{\tiny $\blacktriangleright$}}]}

\DeclareFontFamily{OT1}{rsfso}{}
\DeclareFontShape{OT1}{rsfso}{m}{n}{ <-7> rsfso5 <7-10> rsfso7 <10-> rsfso10}{}
\DeclareMathAlphabet{\myscr}{OT1}{rsfso}{m}{n}

\DeclareMathOperator{\re}{Re}
\DeclareMathOperator{\im}{Im}

\DeclareMathOperator{\Tr}{Tr}
\DeclareMathOperator{\tr}{tr}

\DeclareMathOperator{\supp}{supp}

\DeclareMathOperator{\sign}{sign}
\renewcommand{\u}{\mathfrak{u}}
\renewcommand{\v}{\mathfrak{v}}

\newcommand{\bu}{{\mathbf{u}}}
\newcommand{\bv}{\mathbf{v}}

\newcommand{\bitem}{\begin{itemize}[leftmargin=2.5em]}
\newcommand{\eitem}{\end{itemize}}

\newcommand{\scrt}{T}
\renewcommand{\sc}{\text{\rm{sc}}}

\begin{document}

\maketitle

\begin{abstract}
It is shown for causal fermion systems describing Minkowski-type spacetimes
that an interacting causal fermion system at time $t$ gives rise to a distinguished state
on the algebra generated by fermionic and bosonic field operators.
The proof of positivity of the state is given and representations are constructed.
\end{abstract}

\tableofcontents

\section{Introduction}
The theory of {\em{causal fermion systems}} is a recent approach to fundamental physics
(see the basics in Section~\ref{secprelim}, the reviews~\cite{ nrstg, review, dice2014}, the textbook~\cite{cfs}
or the website~\cite{cfsweblink}).
In this approach, spacetime and all objects therein are described by a measure~$\rho$
on a set~$\F$ of linear operators on a Hilbert space~$(\H, \la .|. \ra_\H)$. 
The physical equations are formulated by means of the so-called {\em{causal action principle}},
a nonlinear variational principle where an action~$\Sact$ is minimized under variations of the measure~$\rho$.

It is an important task to rewrite the dynamics as described by the causal action principle
in a form comparable to the usual description of physics.
The connection to an interaction of a quantized Dirac field via classical bosonic fields
is obtained in the so-called {\em{continuum limit}} as worked out in detail in~\cite{cfs}.
The remaining task is to work out the connection to an interaction via bosonic quantum fields.
A first step in this direction is made in~\cite{qft}, where a Fock space dynamics was
derived starting from the classical field equations obtained in the continuum limit and making
additional assumptions which were physically motivated but not justified mathematically from first principles.
In order to work out the quantum field theory limit of causal fermion systems in a more fundamental
and mathematically more convincing manner, one should not start from the classical field equations
but work instead directly with the Euler-Lagrange (EL) equations corresponding to the causal action
principle and the corresponding conservation laws.
The first step in the resulting research program is the paper~\cite{fockbosonic}, where complex structures and
related Fock space constructions were carried out for the bosonic degrees of freedom.
In the present paper, we include the fermionic degrees of freedom.
Our main result is to show that the causal fermion system at any given time gives rise to a
canonical quantum state which can be represented on the fermionic and bosonic Fock spaces
built up of spinorial wave functions and solutions of the linearized field equations.
Our methods are based on the recent paper~\cite{dirac} where the dynamics of these
spinorial wave functions is studied.
The dynamics of the resulting Fock state will be analyzed in detail in the forthcoming paper~\cite{mix}.
Finally, the research program will be summarized in the survey paper~\cite{qftlimit}.

Our starting point consists of two causal fermion systems~$(\H, \F, \rho)$ (describing the vacuum)
and~$(\tilde{\H}, \tilde{\F}, \tilde{\rho})$ (describing the interacting system).
We assume that both measures~$\rho$ and~$\tilde{\rho}$ are critical points of the causal action.
Our general strategy for constructing the interacting quantum state is to ``compare''
these two systems at a given fixed time and to try to describe the interacting system in terms of linearized fields and
wave functions in the vacuum spacetime. This ``comparison'' can also be understood as a ``measurement''
performed in the interacting spacetime using objects from the vacuum spacetime as ``measurement devices.''
As we shall see, following this strategy will naturally give rise to Fock states having bosonic and fermionic
components.

As a preliminary step, we need to specify what we mean by a ``given fixed time.''
In the vacuum spacetime~$M:=\supp \rho$ we assume a global time function~$T$
(for example the time of an observer in Minkowski space).
In order to obtain a global time function also in the interacting spacetime~$\tilde{M}:=\supp \tilde{\rho}$,
we need to identify the two spacetimes by a mapping
\[ F : M \rightarrow \tilde{M} \:. \]
Clearly, this identification is not unique. This arbitrariness was studied in detail in~\cite{fockbosonic},
and we can treat it here in the same way.

There is another arbitrariness not considered in~\cite{fockbosonic} which is tied to
the fermionic degrees of freedom: For a ``comparison'' of the two systems one needs to identify
the Hilbert spaces~$\H$ and~$\tilde{\H}$. This can be done by choosing a unitary
mapping~$V : \H \rightarrow \tilde{\H}$, but this mapping is determined only up to a unitary transformation.
We thus have the freedom to transform~$V$ according to
\beq \label{arbitraryU}
V \rightarrow V \scrU \qquad \text{with} \qquad \scrU \in \U(\H)\:.
\eeq
This freedom has to be taken into account when ``comparing'' the two systems. It plays a crucial role
in our analysis and will be responsible for the occurrence of quantum states and effects like
quantum entanglement.
More specifically, the analysis in~\cite{fockbosonic} revealed that, for the bosonic degrees of freedom,
the Fock norm can be associated to the exponential of a nonlinear surface layer
integral~$\gamma^t(\tilde{\rho}, \rho)$ (for basics and the notation we refer to Section~\ref{secprelim}).
In order to incorporate the fermionic degrees of freedom, we again use this association, but
treat the arbitrariness~\eqref{arbitraryU} by integrating over~$\scrU$. This leads us to introducing the
\beq \label{Zdef}
\text{partition function} \qquad Z =
\int_\G \exp \Big( \beta \,\gamma^t \big(\tilde{\rho}, \scrU \rho \big) \Big) \: d\mu_\G(\scrU) \:,
\eeq
where~$\G$ is a subgroup of the unitary group with Haar measure~$\mu_\G$, $\beta$ is a real parameter,
and~$\scrU \rho$ is the
unitarily transformed measure defined by~$(\scrU \rho)(\Omega) = \rho (\scrU^{-1} \Omega \scrU)$
(for details see Section~\ref{secZ}). The name ``partition function'' is inspired by the similarity to
statistical physics. The formula~\eqref{Zdef} also resembles the path integral formulation of quantum
field theory (see for example~\cite{pokorski, glimm+jaffe}). However, this similarity is only on a formal level, because in~\eqref{Zdef} we do not integrate
over field configurations, but over the unitary transformations arising in the identification of Hilbert spaces~\eqref{arbitraryU}.

Again in analogy to the procedure in the path integral formulation of quantum field theory, we introduce
the $n$-point functions defining the quantum state~$\omega^t$ by insertions, i.e.\ symbolically
\beq \label{stateintro}
\omega^t(\cdots) := \frac{1}{Z}\: \int_\G (\cdots ) \:\exp \Big( \beta \,\gamma^t \big(\tilde{\rho}, \scrU \rho \big) \Big) \: d\mu_\G(\scrU) \:,
\eeq
where the dots on the left side stand for field operators, whereas the dots on the right are suitable
surface layer integrals (for details see Section~\ref{secqf}).
The insertions can be understood as ``measurements.'' The fact that we integrate over~$\scrU$
at the very end corresponds to the fact that the observer must choose~$\scrU$ before performing
a multi-particle measurement. The $\scrU$-dependence of the insertions gives rise to correlations
between them. These correlations are crucial for getting an entangled state.

It is a major objective of this paper to work out the detailed form of the insertions
and to prove that~\eqref{stateintro} has the positivity properties required for a quantum state.
Once this has been shown, one can also construct representations of the algebra and
rewrite the state~$\omega^t$ as the expectation value of a density operator~$\sigma^t$
on a Fock space~$\Fock$, i.e.\
\[ \omega^t(\cdots) = \Tr_{\Fock} \big( \sigma^t \, \cdots \big) \:. \]
Our analysis of the insertions explains in particular why the fermionic multi-particle wave functions are totally
anti-symmetric. In this way, the Pauli Exclusion Principle is established for causal fermion systems
on the level of totally anti-symmetric wave functions and anti-commuting field operators.

The paper is organized as follows. Section~\ref{secprelim} provides the necessary preliminaries
on causal fermion systems and the causal action principle. We also recall the
definition of various surface layer integrals and summarize the main results from~\cite{dirac} on the
dynamics of spinorial wave functions.
In Section~\ref{secZ} the arbitrariness of the identification of Hilbert spaces~\eqref{arbitraryU} is
explained and built into the nonlinear surface layer integral, leading us to the definition
of the partition function~\eqref{Zdef}. In Section~\ref{secqf} the quantum state at time~$t$ is introduced,
its positivity properties are proven and representations are constructed.
In Section~\ref{secoutlook} we conclude the paper by a brief discussion of the 
algebra of interacting field and the dynamics of our quantum state.
Moreover, we explain a few modifications and refinements of our constructions which may be of relevance
for future developments.
The detailed analysis of the quantum dynamics will be the objective of the follow-up paper~\cite{mix}.

\section{Preliminaries} \label{secprelim}
\subsection{Causal Fermion Systems and the Causal Action Principle}
We now recall the basic setup and introduce the main objects to be used later on.
\begin{Def} \label{defcfs} (causal fermion systems) {\em{ 
Given a separable complex Hilbert space~$\H$ with scalar product~$\la .|. \ra_\H$
and a parameter~$n \in \N$ (the {\em{``spin dimension''}}), we let~$\F \subset \Lin(\H)$ be the set of all
selfadjoint operators on~$\H$ of finite rank, which (counting multiplicities) have
at most~$n$ positive and at most~$n$ negative eigenvalues. On~$\F$ we are given
a positive measure~$\rho$ (defined on a $\sigma$-algebra of subsets of~$\F$).
We refer to~$(\H, \F, \rho)$ as a {\em{causal fermion system}}.
}}
\end{Def} \noindent
A causal fermion system describes a spacetime together
with all structures and objects therein.
In order to single out the physically admissible
causal fermion systems, one must formulate physical equations. To this end, we impose that
the measure~$\rho$ should be a minimizer of the causal action principle,
which we now introduce. For any~$x, y \in \F$, the product~$x y$ is an operator of rank at most~$2n$. 
However, in general it is no longer a selfadjoint operator because~$(xy)^* = yx$,
and this is different from~$xy$ unless~$x$ and~$y$ commute.
As a consequence, the eigenvalues of the operator~$xy$ are in general complex.
We denote these eigenvalues counting algebraic multiplicities
by~$\lambda^{xy}_1, \ldots, \lambda^{xy}_{2n} \in \C$
(more specifically,
denoting the rank of~$xy$ by~$k \leq 2n$, we choose~$\lambda^{xy}_1, \ldots, \lambda^{xy}_{k}$ as all
the non-zero eigenvalues and set~$\lambda^{xy}_{k+1}, \ldots, \lambda^{xy}_{2n}=0$).
We introduce the Lagrangian and the causal action by
\begin{align}
\text{\em{Lagrangian:}} && \L(x,y) &= \frac{1}{4n} \sum_{i,j=1}^{2n} \Big( \big|\lambda^{xy}_i \big|
- \big|\lambda^{xy}_j \big| \Big)^2 \label{Lagrange} \\
\text{\em{causal action:}} && \Sact(\rho) &= \iint_{\F \times \F} \L(x,y)\: d\rho(x)\, d\rho(y) \:. \label{Sdef}
\end{align}
The {\em{causal action principle}} is to minimize~$\Sact$ by varying the measure~$\rho$
under the following constraints,
\begin{align}
\text{\em{volume constraint:}} && \rho(\F) = \text{const} \quad\;\; & \label{volconstraint} \\
\text{\em{trace constraint:}} && \int_\F \tr(x)\: d\rho(x) = \text{const}& \label{trconstraint} \\
\text{\em{boundedness constraint:}} && \iint_{\F \times \F} 
|xy|^2
\: d\rho(x)\, d\rho(y) &\leq C \:, \label{Tdef}
\end{align}
where~$C$ is a given parameter, $\tr$ denotes the trace of a linear operator on~$\H$, and
the absolute value of~$xy$ is the so-called spectral weight,
\beq \label{sw}
|xy| := \sum_{j=1}^{2n} \big|\lambda^{xy}_j \big| \:.
\eeq
This variational principle is mathematically well-posed if~$\H$ is finite-dimensional.
For the existence theory and the analysis of general properties of minimizing measures
we refer to~\cite{discrete, continuum, lagrange}.
In the existence theory one varies in the class of regular Borel measures
(with respect to the topology on~$\Lin(\H)$ induced by the operator norm),
and the minimizing measure is again in this class. With this in mind, here we always assume that
\[ 
\text{$\rho$ is a regular Borel measure}\:. \]

\subsection{Spacetime and Physical Wave Functions}
Let~$\rho$ be a {\em{minimizing}} measure. {\em{Spacetime}}
is defined as the support of this measure,
\[ 
M := \supp \rho \:. \]
Thus the spacetime points are selfadjoint linear operators on~$\H$.
On~$M$ we consider the topology induced by~$\F$ (generated by the operator norm
on~$\Lin(\H)$). Moreover, the measure~$\rho|_M$ restricted to~$M$ gives a volume
measure on spacetime. This makes spacetime into a {\em{topological measure space}}.

The operators in~$M$ contain a lot of information which, if interpreted correctly,
gives rise to spacetime structures like causal and metric structures, spinors
and interacting fields (for details see~\cite[Chapter~1]{cfs}).
Here we restrict attention to those structures needed in what follows.
We begin with the following notion of causality:

\begin{Def} (causal structure) \label{def2}
{\em{ For any~$x, y \in \F$, the product~$x y$ is an operator
of rank at most~$2n$. We denote its non-trivial eigenvalues (counting algebraic multiplicities)
by~$\lambda^{xy}_1, \ldots, \lambda^{xy}_{2n}$.
The points~$x$ and~$y$ are
called {\em{spacelike}} separated if all the~$\lambda^{xy}_j$ have the same absolute value.
They are said to be {\em{timelike}} separated if the~$\lambda^{xy}_j$ are all real and do not all 
have the same absolute value.
In all other cases (i.e.\ if the~$\lambda^{xy}_j$ are not all real and do not all 
have the same absolute value),
the points~$x$ and~$y$ are said to be {\em{lightlike}} separated. }}
\end{Def} \noindent
Restricting the causal structure of~$\F$ to~$M$, we get causal relations in spacetime.

Next, for every~$x \in \F$ we define the {\em{spin space}}~$S_x$ by~$S_x = x(\H)$;
it is a subspace of~$\H$ of dimension at most~$2n$.
It is endowed with the {\em{spin inner product}} $\Sl .|. \Sr_x$ defined by
\[ 
\Sl u | v \Sr_x = -\la u | x v \ra_\H \qquad \text{(for all $u,v \in S_x$)}\:. \]
A {\em{wave function}}~$\psi$ is defined as a function
which to every~$x \in M$ associates a vector of the corresponding spin space,
\[ 
\psi \::\: M \rightarrow \H \qquad \text{with} \qquad \psi(x) \in S_xM \quad \text{for all~$x \in M$}\:. \]
In order to introduce a notion of continuity of a wave function,
we need to compare the wave function at different spacetime points.
Noting that the natural norm on the spin space~$(S_x, \Sl .|. \Sr_x)$ is given by
\[ \big| \psi(x) \big|_x^2 := \big\la \psi(x) \,\big|\, |x|\, \psi(x) \big\ra_\H = \Big\| \sqrt{|x|} \,\psi(x) \Big\|_\H^2 \]
(where~$|x|$ is the absolute value of the symmetric operator~$x$ on~$\H$, and~$\sqrt{|x|}$
is the square root thereof), we say that the wave function~$\psi$ is {\em{continuous}} at~$x$ if
for every~$\varepsilon>0$ there is~$\delta>0$ such that
\beq \label{wavecontinuous}
\big\| \sqrt{|y|} \,\psi(y) -  \sqrt{|x|}\, \psi(x) \big\|_\H < \varepsilon
\qquad \text{for all~$y \in M$ with~$\|y-x\| \leq \delta$} \:.
\eeq
Likewise, $\psi$ is said to be continuous on~$M$ if it is continuous at every~$x \in M$.
We denote the set of continuous wave functions by~$C^0(M, SM)$.

It is an important observation that every vector~$u \in \H$ of the Hilbert space gives rise to a unique
wave function. To obtain this wave function, denoted by~$\psi^u$, we simply project the vector~$u$
to the corresponding spin spaces,
\[ 
\psi^u \::\: M \rightarrow \H\:,\qquad \psi^u(x) = \pi_x u \in S_xM \:. \]
We refer to~$\psi^u$ as the {\em{physical wave function}} of~$u \in \H$.
A direct computation shows that the physical wave functions are continuous
(in the sense~\eqref{wavecontinuous}). Associating to every vector~$u \in \H$
the corresponding physical wave function gives rise to the {\em{wave evaluation operator}}
\[ 
\Psi \::\: \H \rightarrow C^0(M, SM)\:, \qquad u \mapsto \psi^u \:. \]
Every~$x \in M$ can be written as (for the derivation see~\cite[Lemma~1.1.3]{cfs})
\beq
x = - \Psi(x)^* \,\Psi(x) \label{Fid} \:.
\eeq
In words, every spacetime point operator is the local correlation operator of the wave evaluation operator
at this point. This formula is very useful when varying the system, as will be explained in Section~\ref{secELCFS} below.

\subsection{Connection to the Setting of Causal Variational Principles} \label{seccfscvp}
For the analysis of the causal action principle it is most convenient to get into the
simpler setting of causal variational principles. In this setting, $\F$ is a (possibly non-compact)
smooth manifold of dimension~$m \geq 1$ and~$\rho$ a positive Borel measure on~$\F$.
Moreover, we are given a non-negative function~$\L : \F \times \F \rightarrow \R^+_0$
(the {\em{Lagrangian}}) with the following properties:
\bitem
\item[(i)] $\L$ is symmetric: $\L(x,y) = \L(y,x)$ for all~$x,y \in \F$.\label{Cond1}
\item[(ii)] $\L$ is lower semi-continuous, i.e.\ for all \label{Cond2}
sequences~$x_n \rightarrow x$ and~$y_{n'} \rightarrow y$,
\[ \L(x,y) \leq \liminf_{n,n' \rightarrow \infty} \L(x_n, y_{n'})\:. \]
\eitem
The {\em{causal variational principle}} is to minimize the action
\beq \label{Sact} 
\Sact (\rho) = \int_\F d\rho(x) \int_\F d\rho(y)\: \L(x,y) 
\eeq
under variations of the measure~$\rho$, keeping the total volume~$\rho(\F)$ fixed
({\em{volume constraint}}).
If the total volume~$\rho(\F)$ is finite, one minimizes~\eqref{Sact}
over all regular Borel measures with the same total volume.
If the total volume~$\rho(\F)$ is infinite, however, it is not obvious how to implement the volume constraint,
making it necessary to proceed as follows.
We need the following additional assumptions:
\bitem
\item[(iii)] The measure~$\rho$ is {\em{locally finite}}
(meaning that any~$x \in \F$ has an open neighborhood~$U$ with~$\rho(U)< \infty$).\label{Cond3}
\item[(iv)] The function~$\L(x,.)$ is $\rho$-integrable for all~$x \in \F$, giving
a lower semi-continuous and bounded function on~$\F$. \label{Cond4}
\eitem
Given a regular Borel measure~$\rho$ on~$\F$, we then vary over all
regular Borel measures~$\tilde{\rho}$ with
\[ 
\big| \tilde{\rho} - \rho \big|(\F) < \infty \qquad \text{and} \qquad
\big( \tilde{\rho} - \rho \big) (\F) = 0 \]
(where~$|.|$ denotes the total variation of a measure).
These variations of the causal action are well-defined.
The existence theory for minimizers is developed in~\cite{noncompact}.

There are several ways to get from the causal action principle to causal variational principles,
as we now recall. If the Hilbert space~$\H$ is {\em{finite-dimensional}} and the total volume~$\rho(\F)$
is finite, one can proceed as follows:
As a consequence of the trace constraint~\eqref{trconstraint}, for any minimizing measure~$\rho$
the local trace is constant in spacetime, i.e.\
there is a real constant~$c \neq 0$ such that (see~\cite[Theorem~1.3]{lagrange} or~\cite[Proposition~1.4.1]{cfs})
\[ 
\tr x = c \qquad \text{for all~$x \in M$} \:. \]
Restricting attention to operators with fixed trace, the trace constraint~\eqref{trconstraint}
is equivalent to the volume constraint~\eqref{volconstraint} and may be disregarded.
The boundedness constraint, on the other hand, can be treated with a Lagrange multiplier.
More precisely, in~\cite[Theorem~1.3]{lagrange} it is shown that for every minimizing measure~$\rho$, 
there is a Lagrange multiplier~$\kappa>0$ such that~$\rho$ is a critical point of the causal action
with the Lagrangian replaced by
\beq \label{Lkappa}
\L_\kappa(x,y) := \L(x,y) + \kappa\, |xy|^2 \:,
\eeq
leaving out the boundedness constraint.
Having treated the constraints, the difference to causal variational principles is that
in the setting of causal fermion systems, the set of operators~$\F \subset \Lin(\H)$ does not have the
structure of a manifold. In order to give this set a manifold structure,
we assume that a given minimizing measure~$\rho$ (for the Lagrangian~$\L_\kappa$)
is {\em{regular}} in the sense that all operators in its support
have exactly~$n$ positive and exactly~$n$ negative eigenvalues.
This leads us to introduce the set~$\F^\reg$ 
as the set of all operators~$F$ on~$\H$ with the following properties:
\begin{itemize}[leftmargin=2em]
\item[(i)] $F$ is selfadjoint, has finite rank and (counting multiplicities) has
exactly~$n$ positive and~$n$ negative eigenvalues. \\[-0.8em]
\item[(ii)] The trace is constant, i.e
\beq \tr(F) = c>0 \:. \label{trconst2}
\eeq
\end{itemize}
The set~$\F^\reg$ has a smooth manifold structure
(see the concept of a flag manifold in~\cite{helgason} or the detailed construction
in~\cite[Section~3]{gaugefix}). In this way, the causal action principle becomes an
example of a causal variational principle.

This finite-dimensional setting has the drawback that the total volume~$\rho(\F)$ of spacetime
is finite, which is not suitable for describing asymptotically flat spacetimes or spacetimes of
infinite lifetime like Minkowski space. Therefore, it is important to also consider
the {\em{infinite-dimensional setting}} where~$\dim \H=\infty$ and consequently also~$\rho(\F) = \infty$ 
(see~\cite[Exercise~1.3]{cfs}). In this case, the set~$\F^\reg$
has the structure of an infinite-dimensional Banach manifold (for details see~\cite{banach}).
Here we shall not enter the subtleties of infinite-dimensional analysis. Instead,
we get by with the following simple method: Given a minimizing measure~$\rho$,
we choose~$\F^\reg$ as a finite-dimensional manifold which contains~$M:=\supp \rho$.
We then restrict attention to variations of~$\rho$ in the class of regular Borel measures on~$\F^\reg$.
In this way, we again get into the setting of causal variational principles.
We refer to this method by saying that we {\em{restrict attention to locally compact variations}}.
Keeping in mind that the dimension of~$\F^\reg$ can be chosen arbitrarily large, this
method seems a sensible technical simplification. In situations when it is important to
work in infinite dimensions (for example for getting the connection
to the renormalization program in quantum field theory), it may be necessary to analyze
the limit when the dimension of~$\F^\reg$ tends to infinity, or alternatively it may be suitable to work in
the infinite-dimensional setting as developed in~\cite{banach}. However, this is not a concern of the present paper,
where we try to keep the mathematical setup as simple as possible.

For ease in notation, in what follows we will omit the superscript ``$\reg$.'' Thus~$\F$ stands
for a smooth (in general non-compact) manifold which contains the support~$M$ of a given minimizing
measure~$\rho$.

\subsection{The Euler-Lagrange Equations and Jet Spaces} \label{secEL}
A minimizer of a causal variational principle
satisfies the following {\em{Euler-Lagrange (EL) equations}}.
For a suitable value of the parameter~$\s>0$,
the lower semi-continuous function~$\ell : \F \rightarrow \R_0^+$ defined by
\[ 
\ell(x) := \int_M \L(x,y)\: d\rho(y) - \s \]
is minimal and vanishes on spacetime~$M:= \supp \rho$,
\beq \label{EL}
\ell|_M \equiv \inf_\F \ell = 0 \:.
\eeq
The parameter~$\s$ can be understood as the Lagrange parameter
corresponding to the volume constraint. For the derivation and further details we refer to~\cite[Section~2]{jet}.

The EL equations~\eqref{EL} are nonlocal in the sense that
they make a statement on the function~$\ell$ even for points~$x \in \F$ which
are far away from spacetime~$M$.
It turns out that for the applications we have in mind, it is preferable to
evaluate the EL equations only locally in a neighborhood of~$M$.
This leads to the {\em{weak EL equations}} introduced in~\cite[Section~4]{jet}.
Here we give a slightly less general version of these equations which
is sufficient for our purposes. In order to explain how the weak EL equations come about,
we begin with the simplified situation that the function~$\ell$ is smooth.
In this case, the minimality of~$\ell$ implies that the derivative of~$\ell$
vanishes on~$M$, i.e.\
\beq \label{ELweak}
\ell|_M \equiv 0 \qquad \text{and} \qquad D \ell|_M \equiv 0
\eeq
(where~$D \ell(p) : T_p \F \rightarrow \R$ is the derivative).
In order to combine these two equations in a compact form,
it is convenient to consider a pair~$\u := (a, \bu)$
consisting of a real-valued function~$a$ on~$M$ and a vector field~$\bu$
on~$T\F$ along~$M$, and to denote the combination of 
multiplication of directional derivative by
\beq \label{Djet}
\nabla_{\u} \ell(x) := a(x)\, \ell(x) + \big(D_\bu \ell \big)(x) \:.
\eeq
Then the equations~\eqref{ELweak} imply that~$\nabla_{\u} \ell(x)$
vanishes for all~$x \in M$.
The pair~$\u=(a,\bu)$ is referred to as a {\em{jet}}.

In the general lower-continuous setting, one must be careful because
the directional derivative~$D_\bu \ell$ in~\eqref{Djet} need not exist.
Our method for dealing with this problem is to restrict attention to vector fields
for which the directional derivative is well-defined.
Moreover, we must specify the regularity assumptions on~$a$ and~$u$.
To begin with, we always assume that~$a$ and~$\bu$ are {\em{smooth}} in the sense that they
have a smooth extension to the manifold~$\F$ (for more details see~\cite[Section~2.2]{fockbosonic}).
Thus the jet~$\u$ should be
an element of the jet space
\[ \J_\rho := \big\{ \u = (a,\bu) \text{ with } a \in C^\infty(M, \R) \text{ and } \bu \in \Gamma(M, T\F) \big\} \:, \]
where~$C^\infty(M, \R)$ and~$\Gamma(M,T\F)$ denote the space of real-valued functions and vector fields
on~$M$, respectively, which admit a smooth extension to~$\F$.

Clearly, the fact that a jet~$\u$ is smooth does not imply that the functions~$\ell$
or~$\L$ are differentiable in the direction of~$\u$. This must be ensured by additional
conditions which are satisfied by suitable subspaces of~$\J_\rho$
which we now introduce.
First, we let~$\Gdiff_\rho$ be those vector fields for which the
directional derivative of the function~$\ell$ exists,
\[ \Gdiff_\rho = \big\{ \bu \in C^\infty(M, T\F) \;\big|\; \text{$D_{\bu} \ell(x)$ exists for all~$x \in M$} \big\} \:. \]
This gives rise to the jet space
\[ \Jdiff_\rho := C^\infty(M, \R) \oplus \Gdiff_\rho \;\subset\; \J_\rho \:. \]
For the jets in~$\Jdiff_\rho$, the combination of multiplication and directional derivative
in~\eqref{Djet} is well-defined. 
We choose a linear subspace~$\Jtest_\rho \subset \Jdiff_\rho$ with the property
that its scalar and vector components are both vector spaces,
\[ \Jtest_\rho = \Ctest(M, \R) \oplus \Gtest_\rho \;\subseteq\; \Jdiff_\rho \:, \]
and the scalar component is nowhere trivial in the sense that
\beq \label{Cnontriv}
\text{for all~$x \in M$ there is~$a \in \Ctest(M, \R)$ with~$a(x) \neq 0$}\:.
\eeq
Then the {\em{weak EL equations}} read (for details cf.~\cite[(eq.~(4.10)]{jet})
\beq \label{ELtest}
\nabla_{\u} \ell|_M = 0 \qquad \text{for all~$\u \in \Jtest_\rho$}\:.
\eeq
Before going on, we point out that the weak EL equations~\eqref{ELtest}
do not hold only for minimizers, but also for critical points of
the causal action. With this in mind, all methods and results of this paper 
do not apply only to
minimizers, but more generally to critical points of the causal variational principle.
For brevity, we also refer to a measure which satisfies the weak EL equations~\eqref{ELtest}
as a {\em{critical measure}}.

Here and throughout this paper, we use the following conventions for partial derivatives and jet derivatives:
\bitem
\itemD Partial and jet derivatives with an index~$i \in \{ 1,2 \}$ only act on the respective variable of the function $\L$.
This implies, for example, that the derivatives commute,
\[ 
\nabla_{1,\v} \nabla_{1,\u} \L(x,y) = \nabla_{1,\u} \nabla_{1,\v} \L(x,y) \:. \]
\itemD The partial or jet derivatives which do not carry an index act as partial derivatives
on the corresponding argument of the Lagrangian. This implies, for example, that
\[ \nabla_\u \int_\F \nabla_{1,\v} \, \L(x,y) \: d\rho(y) =  \int_\F \nabla_{1,\u} \nabla_{1,\v}\, \L(x,y) \: d\rho(y) \:, \]
where the notation~$\nabla_\u$ without an index~$1$ or~$2$ means that the derivative acts on the
only free variable~$x$.
\eitem
We point out that, with these conventions, {\em{jets are never differentiated}}.
Finally, compactly supported jets are denoted by a subscript zero, like for example
\[ 
\Jtest_{\rho,0} := \{ \u \in \Jtest_\rho \:|\: \text{$\u$ has compact support} \} \:. \]

\subsection{The Euler-Lagrange Equations for the Physical Wave Functions} \label{secELCFS}
For causal fermion systems, the EL equations can be expressed in terms
of the physical wave functions, as we now recall. These equations were first derived
in~\cite[\S1.4.1]{cfs} (based on a weaker version in~\cite[Section~3.5]{pfp}),
even before the jet formalism was developed. We now make the connection between the
different formulations, in a way most convenient for our constructions.
Our starting point is the formula~\eqref{Fid} which expresses the spacetime point operator
as a local correlation operator. Varying the wave evaluation operator gives a vector field~$\bu$ on~$\F$
along~$M$,
\beq \label{ufermi}
\bu(x) = -\delta \Psi(x)^*\, \Psi(x) - \Psi(x)^*\, \delta \Psi(x) \:.
\eeq
In order to make mathematical sense of this formula in agreement with the concept of restricting
attention to locally compact variations, we choose a finite-dimensional subspace~$\H^\fermi \subset \H$, i.e.\
\beq \label{Hfermidef}
f^\fermi := \dim \H^\fermi  < \infty
\eeq
and impose the following assumptions on~$\delta \Psi$
(similar variations were first considered in~\cite[Section~7]{perturb}):
\begin{itemize}[leftmargin=2em]
\item[\rm{(a)}] 
The variation is trivial on the orthogonal complement of~$\H^\fermi$,
\[ \delta \Psi |_{(\H^\fermi)^\perp} = 0 \:. \]
\item[\rm{(b)}] The variations of all physical wave functions are continuous and compactly supported, i.e.
\[ \delta \Psi : \H \rightarrow C^0_0(M, SM) \:. \]
\end{itemize}
Before going on, we point out that the choice of~$\H^\fermi$ is {\em{not canonical}}.
Ultimately, one would like to exhaust~$\H$ by a sequence of finite-dimensional subspaces~$\H^\fermi_1 \subset
\H^\fermi_2 \subset \cdots$ and take the limit. Here we shall not enter this analysis, but 
for technical simplicity we rather choose~$\H^\fermi$ as a finite-dimensional subspace of sufficiently large dimension.

We choose~$\Gfermi_{\rho,0}$ as a space of vector fields of the form~\eqref{ufermi}.
For convenience, we identify the vector field with the first variation~$\delta \Psi$ and
write~$\delta \Psi \in \Gfermi_{\rho,0}$ (this representation of~$\bu$ in terms of~$\delta \Psi$
may not be unique, but this is of no relevance for what follows).
Choosing trivial scalar components, we obtain a corresponding space of jets~$\Jfermi_{\rho, 0}$,
referred to as the {\em{fermionic jets}}. We always assume that the fermionic jets are
admissible for testing, i.e.\
\[ \J^\fermi_{\rho,0} := \{0\} \oplus \Gfermi_{\rho,0} \;\subset\; \Jtest_{\rho,0} \:. \]
Moreover, in analogy to the condition~\eqref{Cnontriv} for the scalar components of the test jets,
we assume that the variation can have arbitrary values at any spacetime point, i.e.\
\beq \label{Gnontriv}
\text{for all~$x \in M, \chi \in S_x$ and~$\phi \in \H^\fermi$ there is~$\delta \Psi \in \Gfermi_{\rho,0}$ with~$\delta \Psi(x)\, \phi = \chi$}\:.
\eeq

For the computation of the variation of the Lagrangian, one can make use of the fact
that for any $p \times q$-matrix~$A$ and any~$q \times p$-matrix~$B$,
the matrix products~$AB$ and~$BA$ have the same non-zero eigenvalues, with the same
algebraic multiplicities. As a consequence, applying again~\eqref{Fid},
\beq
x y 
= \Psi(x)^* \,\big( \Psi(x)\, \Psi(y)^* \Psi(y) \big)
\simeq \big( \Psi(x)\, \Psi(y)^* \Psi(y) \big)\,\Psi(x)^* \:, \label{isospectral}
\eeq
where $\simeq$ means that the operators have the same non-trivial eigenvalues
with the same algebraic multiplicities. Introducing the {\em{kernel of the fermionic projector}} $P(x,y)$ by
\[ P(x,y) := \Psi(x)\, \Psi(y)^* \::\: S_y \rightarrow S_x \:, \]
we can write~\eqref{isospectral} as
\[ x y \simeq P(x,y)\, P(y,x) \::\: S_x \rightarrow S_x \:. \]
In this way, the eigenvalues of the operator product~$xy$ as needed for the computation of
the Lagrangian~\eqref{Lagrange} and the spectral weight~\eqref{sw} are recovered as
the eigenvalues of a $2n \times 2n$-matrix. Since~$P(y,x) = P(x,y)^*$,
the Lagrangian~$\L_\kappa(x,y)$ in~\eqref{Lkappa} can be formulated in terms of~$P(x,y)$.
Consequently, also the first variation of the Lagrangian can be expressed in terms
of the first variation of this kernel. Being real-valued and real-linear in~$\delta P(x,y)$,
it can be written as
\beq \label{delLdef}
\delta \L_\kappa(x,y) = 2 \re \Tr_{S_x} \!\big( Q(x,y)\, \delta P(x,y)^* \big)
\eeq
with a kernel~$Q(x,y)$ which is again symmetric (with respect to the spin inner product), i.e.
\[ Q(x,y) \::\: S_y \rightarrow S_x \qquad \text{and} \qquad Q(x,y)^* = Q(y,x) \]
(more details on this method and many computations can be found in~\cite[Sections~1.4 and~2.6
as well as Chapters~3-5]{cfs}). Expressing the variation of~$P(x,y)$ in terms of~$\delta \Psi$,
the variations of the Lagrangian can be written as
\begin{align*}
D_{1,\bu} \L_\kappa(x,y) = 2\, \re \tr \big( \delta \Psi(x)^* \, Q(x,y)\, \Psi(y) \big) \\
D_{2,\bu} \L_\kappa(x,y) = 2\, \re \tr \big( \Psi(x)^* \, Q(x,y)\, \delta \Psi(y) \big)
\end{align*}
(where~$\tr$ denotes the trace of a finite-rank operator on~$\H$). Likewise, the
variation of~$\ell$ becomes
\[ D_\bu \ell(x) = 2\, \re \int_M \tr \big( \delta \Psi(x)^* \, Q(x,y)\, \Psi(y) \big)\: d\rho(y) \:. \]
The weak EL equations~\eqref{ELtest} imply that this expression must vanish for any~$\bu \in \Gfermi_{\rho,0}$.
Using that the variation can be arbitrary at every spacetime point (see~\eqref{Gnontriv}), one may be tempted
to conclude that
\[ 
\int_M Q(x,y)\, \Psi(y) \:\phi\: d\rho(y) = 0 \qquad \text{for all~$x \in M$ and~$\phi \in \H^\fermi$}\:. \]
However, we must take into account that the local trace must be
preserved in the variation~\eqref{trconst2}. This can be arranged by rescaling the
operator~$x$ in the variation (for details see~\cite[Section~6.2]{perturb}) or, equivalently,
by treating it with a Lagrange multiplier term (see~\cite[\S1.4.1]{cfs}). We thus obtain
the {\em{EL equation for the physical wave functions}}
\beq \label{ELQ}
\int_M Q(x,y)\, \Psi^\fermi(y) \:d\rho(y) = \mathfrak{r}\, \Psi^\fermi(x) \qquad \text{for all~$x \in M$}\:,
\eeq
where~$\mathfrak{r} \in \R$ is the Lagrange parameter of the trace constraint,
and~$\Psi^\fermi:=\Psi|_{\H^\fermi}$ denotes the restriction of the wave evaluation operator
to the finite-dimensional subspace~$\H^\fermi$.

We finally comment on the significance of the subspace~$\H^\fermi \subset \H$.
Choosing a finite-dimensional subspace is a technical simplification, made in agreement with the
method of restricting attention to locally compact variations discussed in Section~\ref{seccfscvp}.
The strategy is to choose~$\H^\fermi$ so large that all relevant physical effects are captured.
The general picture is that~$\H^\fermi$ should contain all physical wave functions whose energies
are much smaller than the Planck energy.
If necessary, one could analyze the limit where the dimension~$f^\fermi$ of~$\H^\fermi$ tends to infinity.
In what follows, we leave~$\H^\fermi$ unspecified and merely assume it to be
a finite-dimensional subspace of~$\H$.

\subsection{Surface Layer Integrals for Jets} \label{secosi}
{\em{Surface layer integrals}} were first introduced in~\cite{noether}
as double integrals of the general form
\beq \label{osi}
\int_\Omega \bigg( \int_{M \setminus \Omega} (\cdots)\: \L_\kappa(x,y)\: d\rho(y) \bigg)\, d\rho(x) \:,
\eeq
where $(\cdots)$ stands for a suitable
differential operator formed of jets.
A surface layer integral generalizes the concept of a surface integral over~$\partial \Omega$
to the setting of causal fermion systems.
The connection can be understood most easily in the
case when~$\L_\kappa(x,y)$ vanishes
unless~$x$ and~$y$ are close together. In this case, we only get a contribution to~\eqref{osi}
if both~$x$ and~$y$ are close to the boundary of~$\Omega$.
A more detailed explanation of the idea of a surface layer integral is given in~\cite[Section~2.3]{noether}.

We now recall those surface layer integrals for jets which will be of relevance in this paper.
\begin{Def} \label{defosi} We define the following surface layer integrals,
\begin{align}
\gamma^\Omega_\rho \::\: \J_{\rho, \sc} &\rightarrow \R \qquad \text{(conserved one-form)} \notag \\
\gamma^\Omega_\rho(\v) &= \int_{\Omega} d\rho(x) \int_{M \setminus \Omega} d\rho(y)\:
\big( \nabla_{1,\v} - \nabla_{2,\v} \big) \L(x,y) \label{gamma} \\
\sigma^\Omega_\rho \::\: \J_{\rho, \sc} \times \J_{\rho, \sc} &\rightarrow \R  \qquad \text{(symplectic form)} \notag \\
\sigma^\Omega_\rho(\u, \v) &= \int_{\Omega} d\rho(x) \int_{M \setminus \Omega} d\rho(y)\:
\big( \nabla_{1,\u} \nabla_{2,\v} - \nabla_{2,\u} \nabla_{1,\v} \big) \L(x,y) \label{sigma} \\
(.,.)^\Omega_\rho \::\: \J_{\rho, \sc} \times \J_{\rho, \sc} &\rightarrow \R \qquad \text{(surface layer inner product)} \notag \\
(\u, \v)^\Omega_\rho &= \int_{\Omega} d\rho(x) \int_{M \setminus \Omega} d\rho(y)\:
\big( \nabla_{1,\u} \nabla_{1,\v} - \nabla_{2,\u} \nabla_{2,\v} \big) \L(x,y) \:. \label{sprod}
\end{align}
\end{Def} \noindent
Here~$\J_{\rho, \sc}$ denotes the jets in~$\Jvary_\rho$ with spatially compact support
(for details see~\cite[Section~5.3]{linhyp}),
where~$\Jvary_\rho$ is a suitably chosen subspace of~$\Jtest_\rho$
(for details see~\cite[Section~3.2]{linhyp}).

\subsection{The Linearized Field Equations and Bosonic Green's Operators} \label{seclinfield}
In simple terms, the linearized field equations
describe variations of the measure~$\rho$ which preserve the EL equations.
More precisely, we consider variations where we multiply~$\rho$ by a non-negative
function and take the push-forward with respect to a mapping from~$M$ to~$\F$.
Thus we consider families of measures~$(\tilde{\rho}_\tau)_{\tau \in (-\delta, \delta)}$ 
of the form
\beq \label{rhotau}
\tilde{\rho}_\tau = (F_\tau)_* \big( f_\tau \, \rho \big) \:,
\eeq
where the~$f_\tau$ and~$F_\tau$ are smooth,
\[ f_\tau \in C^\infty\big(M, \R^+ \big) \qquad \text{and} \qquad
F_\tau \in C^\infty\big(M, \F \big) \:, \]
depend smoothly on the parameter~$\tau$
and have the properties~$f_0(x)=1$ and~$F_0(x) = x$ for all~$x \in M$
(moreover, the star denotes the push-forward measure, which is defined
for a subset~$\Omega \subset \F$ by~$((F_\tau)_*\mu)(\Omega)
= \mu ( F_\tau^{-1} (\Omega))$; see for example~\cite[Section~3.6]{bogachev}).
We assume that the measures~$(\tilde{\rho}_\tau)_{\tau \in (-\delta, \delta)}$ satisfy
the EL equations~\eqref{EL} for all~$\tau$, i.e.
\[ 
\tilde{\ell}_\tau|_{M_\tau} \equiv \inf_\F \ell_\tau = 0 
\qquad \text{with} \qquad \tilde{\ell}_\tau(x) := \int_\F \L_\kappa(x,y)\: d\tilde{\rho}_\tau(y) - \s \:. \]
Using the definition of the push-forward measure, we can rewrite the $\tilde{\rho}$-integral
as a $\rho$-integral. Moreover,
it is convenient to rewrite this equation as an equation on~$M$ and to multiply it
by~$f_\tau(x)$. We thus obtain the equivalent equation
\[ \ell_\tau|_M \equiv \inf_\F \ell_\tau = 0 \]
with
\[ \ell_\tau(x) :=
\int_\F f_\tau(x) \,\L_\kappa\big(F_\tau(x),F_\tau(y) \big)\: f_\tau(y)\: d\tilde{\rho}_\tau(y) - f_\tau(x)\: \s\:. \]
In analogy to~\eqref{ELtest} we write the corresponding weak EL equations as
\[ 
\nabla_{\u} \ell_\tau|_M = 0 \qquad \text{for all~$\u \in \Jtest_\rho$} \]
(for details on why the jet space does not depend on~$\tau$ we refer to~\cite[Section~4.1]{perturb}).
Since this equation holds by assumption for all~$\tau$, we can differentiate it with respect to~$\tau$.
Denoting the infinitesimal generator of the variation by~$\v$, i.e.
\[ 
\v(x) := \frac{d}{d\tau} \big( f_\tau(x), F_\tau(x) \big) \Big|_{\tau=0} \:, \]
we thus obtain the {\em{linearized field equations}}
\[ 
0 = \la \u, \Delta \v \ra(x) := 
\nabla_\u \bigg( \int_M \big( \nabla_{1, \v} + \nabla_{2, \v} \big) \L_\kappa(x,y)\: d\rho(y) - \nabla_\v \,\s \bigg) \:, \]
which are to be satisfied for all~$\u \in \Jtest_\rho$ and all~$x \in M$
(for details see~\cite[Section~3.3]{perturb}).
We denote the vector space of all solutions of the linearized field equations by~$\Jlin_\rho$.

In~\cite[Section~5]{linhyp} advanced and retarded Green's operators are constructed as mappings
\[ 
S^\vee, S^\wedge \::\: \J^*_0 \rightarrow L^2_\sc(M, d\rho) \:, \]
where~$J^*_0$ is a space of compactly supported jets, whereas~$L^2_\sc(M, d\rho)$
are locally square-integrable jets with spatially compact support.
They are inverses of the linearized field operator in the sense that~$\Delta S^{\vee, \wedge}=\1$.
The difference of these Green's operators is the so-called causal fundamental solution~$G$;
it maps to linearized solutions,
\beq \label{Jlinscdef}
G := S^\wedge - S^\vee \,:\, \J^*_0  \rightarrow \Jlin_\sc \:.
\eeq
It satisfies for all~$\u, \v \in \J_0^*$ the relation
\beq \label{sigmaform}
\sigma(G\u, G\v) = \la \u, G \,\v \ra_{L^2(M)} \:.
\eeq

\subsection{Minkowski-Type Spacetimes} \label{secminktype}
Let~$(\H, \F, \rho)$ be a causal fermion system, which may be thought of as describing either the vacuum
or the interacting physical system. We assume that~$\rho$ is a critical point of the
causal action principle. Moreover, we assume that
the corresponding spacetime~$M:= \supp \rho$ is diffeomorphic to a four-dimensional spacetime
with trivial topology, i.e.
\[ M \simeq \scrM := \R^4 \:. \]
Next, we assume that, using this identification, the measure~$\rho$ is absolutely continuous with respect
to the Lebesgue measure with a smooth weight function, i.e.\
\beq \label{rhoh}
d\rho = h(x)\: d^4x \qquad \text{with} \quad h \in C^\infty(\scrM, \R^+) \:.
\eeq
We also assume that~$h$ is bounded from above and below, i.e.\ there should be a constant~$C>1$ with
\[ 
\frac{1}{C} \leq h(x) \leq C \qquad \text{for all~$x \in \scrM$} \:. \]

We also denote the coordinate~$x^0$ as time function~$\scrt$,
\beq \label{scrtdef}
\scrt \::\: M \rightarrow \R\:,\qquad (t, \x) \mapsto t \:.
\eeq
For any~$t \in \R$, we let~$\Omega^t$ be the past of~$t$,
\[ 
\Omega^t := \{ x \in \scrM \:|\: \scrt(x) \leq t \} \:. \]

\subsection{Inner Solutions, Arranging Jets without Scalar Components} \label{seczeroscalar}
We now briefly recall the definition of inner solutions as introduced in~\cite[Section~3]{fockbosonic}.

\begin{Def} \label{definner} An {\bf{inner solution}} is a jet~$\v$ of the form
\[ \v = (\div \bv, \bv) \qquad \text{with} \qquad \bv \in \Gamma(M, TM) \:, \]
where the divergence is taken with respect to the measure in~\eqref{rhoh},
\[ \div \bv := \frac{1}{h}\: \partial_j \big( h\, \bv^j \big) \:. \]
\end{Def}
Under suitable regularity and decay assumptions, an inner solution solves the
linearized field equations (for details see~\cite[Section~3.1]{fockbosonic}).
We denote these inner solutions by~$\Jin_\rho$. In~\cite[Proposition~3.6]{fockbosonic}
it is shown that inner solutions can be used for testing. With this in mind, we always assume that
\[ 
\Jin_\rho \subset \Jtest_\rho \:. \]

Inner solutions can be regarded as infinitesimal generators of transformations of~$M$
which leave the measure~$\rho$ unchanged. Therefore, inner solutions do not change the
causal fermion system, but merely describe symmetry transformations of the measure.
With this in mind, we can modify solutions of the linearized field equations by adding
inner solutions. For conceptual clarity, it is preferable that the diffeomorphism
generated by an inner solution does not change the global time function~$T$ in~\eqref{scrtdef}.
Therefore, we consider vector fields which are tangential to the
hypersurfaces~$N_t := T^{-1}(t)$. As shown in~\cite[Lemma~2.7]{pmt}, the divergence of
such a vector field can be arranged to be any given function~$a \in C^\infty(M, \R)$.
Thus, by adding the corresponding inner solutions we can achieve that all linearized solutions
have no scalar components. Therefore, in what follows we may restrict attention to linearized
solutions~$\v \in \Jlin$ with vanishing scalar component. We also write these jets as
\[ \v = (0, \bv) \qquad \text{with} \qquad \bv \in \Glin_\rho \:. \]
Likewise, the causal fundamental solution maps to linearized solutions without scalar
components. Thus~\eqref{Jlinscdef} is modified to
\[ 
G := S^\wedge - S^\vee \,:\, \J^*_0  \rightarrow \Glin_\sc \:. \]

\subsection{The Extended Hilbert Space and Fermionic Green's Operators} \label{secHextend}
The EL equation~\eqref{ELQ} is a linear equation describing
the dynamics of the physical wave functions. In~\cite{dirac}, this wave equation
was extended to more general solutions which form the so-called 
{\em{extended Hilbert space}}~$(\H^{\fermi,t}_\rho, \la .|. \ra^t_\rho)$.
The dynamics in~$\H^{\fermi,t}_\rho$ is described by the so-called {\em{dynamical wave equation}}
\[ 
Q^\dyn \psi = 0 \:, \]
where~$Q^\dyn$ is an integral operator with a symmetric integral kernel
\[ Q^\dyn(x,y) \,:\, S_y \rightarrow S_x \:. \]
The scalar product~$\la .|. \ra^t_\rho$ at time~$t$ has the form
\beq \label{OSIdyn}
\begin{split}
\la \psi | \phi \ra^t_\rho = -2i \,\bigg( \int_{\Omega^t} \!d\rho(x) \int_{M \setminus \Omega^t} \!\!\!\!\!\!\!d\rho(y) 
&- \int_{M \setminus \Omega^t} \!\!\!\!\!\!\!d\rho(x) \int_{\Omega^t} \!d\rho(y) \bigg)\\
&\times\:
\Sl \psi(x) \:|\: Q^\dyn(x,y)\, \phi(y) \Sr_x \:.
\end{split}
\eeq

Moreover, the fermionic {\em{causal fundamental solution}} is introduced as the mapping
\beq \label{kdef}
k:= \frac{i}{2} \big( s^\vee - s^\wedge \big) \::\: \scrW^*_{\mathrm{fpc}}(M,SM) \rightarrow \H^\fermi_\sc(M, SM) \:,
\eeq
where~$s^\vee$ and~$s^\wedge$ are the advanced and retarded Green's operators, respectively,
which map from a space of wave functions~$\scrW^*_{\mathrm{fpc}}(M,SM)$ supported in finite time strips
to a space~$\H^\fermi_\sc(M, SM)$ of spatially compact wave functions (for details see~\cite[Section~6]{dirac}).
The following relation holds for all~$\eta, \eta' \in \scrW_0$,
\beq \label{kkrel}
\la k \eta \,|\, k \eta' \ra^t_\rho = \bra \eta \,|\, k \,\eta' \ket \:,
\eeq
where the last sesquilinear form is the Krein inner product (for details see~\cite[\S1.1.5]{cfs})
\beq \label{krein}
\bra \eta | \eta' \ket := \int_M \Sl \eta(x) \,|\, \eta'(x) \Sr_x\: d\rho(x) \:.
\eeq

\subsection{A Conserved Nonlinear Surface Layer Integral} \label{secosinonlin}
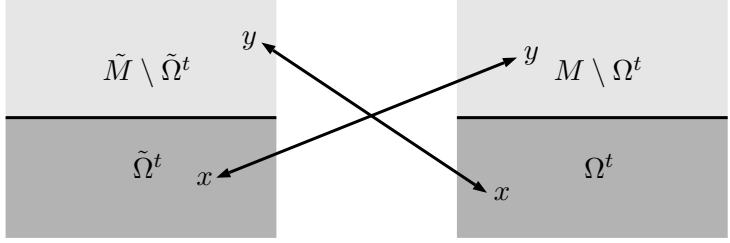
\begin{figure}[tb]
\psscalebox{1.0 1.0} 
{
\begin{pspicture}(0,26.0)(9.62,29.2)
\definecolor{colour0}{rgb}{0.7019608,0.7019608,0.7019608}
\definecolor{colour1}{rgb}{0.9019608,0.9019608,0.9019608}
\psframe[linecolor=colour0, linewidth=0.04, fillstyle=solid,fillcolor=colour0, dimen=outer](9.61,27.6)(6.01,26.0)
\psframe[linecolor=colour1, linewidth=0.04, fillstyle=solid,fillcolor=colour1, dimen=outer](9.61,29.2)(6.01,27.6)
\psframe[linecolor=colour0, linewidth=0.04, fillstyle=solid,fillcolor=colour0, dimen=outer](3.61,27.6)(0.01,26.0)
\psframe[linecolor=colour1, linewidth=0.04, fillstyle=solid,fillcolor=colour1, dimen=outer](3.61,29.2)(0.01,27.6)
\psline[linecolor=black, linewidth=0.04](0.01,27.6)(3.61,27.6)
\psline[linecolor=black, linewidth=0.04](6.01,27.6)(9.61,27.6)
\psline[linecolor=black, linewidth=0.04, arrowsize=0.05291667cm 2.0,arrowlength=1.4,arrowinset=0.0]{<->}(3.41,28.6)(6.41,26.6)
\psline[linecolor=black, linewidth=0.04, arrowsize=0.05291667cm 2.0,arrowlength=1.4,arrowinset=0.0]{<->}(2.81,26.8)(6.81,28.4)
\rput[bl](7.3,28){$M \setminus \Omega^t$}
\rput[bl](7.7,26.8){$\Omega^t$}
\rput[bl](1.3,28){$\tilde{M} \setminus \tilde{\Omega}^t$}
\rput[bl](1.7,26.8){$\tilde{\Omega}^t$}
\rput[bl](2.55,26.7){$x$}
\rput[bl](3.15,28.5){$y$}
\rput[bl](6.5,26.5){$x$}
\rput[bl](6.9,28.3){$y$}
\end{pspicture}
}
\caption{The nonlinear surface layer integral.}
\label{figosinl}
\end{figure}

In~\cite[Section~4]{fockbosonic} a nonlinear surface layer integral~$\gamma^t(\rho, \tilde{\rho})$
was introduced which can be used for comparing the measure~$\tilde{\rho}$ describing the interacting system
with the vacuum measure~$\rho$. As the starting point, we let~$\rho$ and~$\tilde{\rho}$ be
two critical measures  on~$\F$, which describe the
vacuum and the interacting system, respectively. Similar as in~\eqref{rhotau} we assume that
the interacting measure can be written as
\beq \label{tilrho}
\tilde{\rho} = F_* (f \rho)
\eeq
with smooth functions~$f \in C^\infty(M, \R^+)$ and~$F \in C^\infty(M, \F \big)$.
Moreover, we assume that the mapping~$F$ is injective and closed, implying that
\[ \tilde{M} := \supp \tilde{\rho} = F(M) \:, \]
and that the mapping~$F : M \rightarrow \tilde{M}$ is invertible.
We denote the inverse of this mapping by~$\Phi : \tilde{M} \rightarrow M$.
Choosing a foliation~$(N_t)_{t \in \R}$ of~$M$,
we obtain a corresponding foliation~$(\tilde{N}_t)_{t \in \R}$ of~$\tilde{M}$
given by~$\tilde{N}_t = F(N_t)$.
Likewise, past sets~$\Omega^t \subset M$ correspond to past sets~$\tilde{\Omega}^t \subset \tilde{M}$.
Then the nonlinear surface layer integral at time~$t$ is defined by (see
Figure~\ref{figosinl} and more details in~\cite[Definition~4.1]{fockbosonic})

\beq \label{osinl}
\gamma^t(\tilde{\rho}, \rho) =
\int_{\tilde{\Omega}^t} d\tilde{\rho}(x) \int_{M \setminus \Omega^t} d\rho(y)\: \L(x,y)
- \int_{\Omega^t} d\rho(x) \int_{\tilde{M} \setminus \tilde{\Omega}^t}  d\tilde{\rho}(y)\: \L(x,y) \:.
\eeq
Using the definition of the push-forward measure, this can be
written alternatively as
\beq \label{osinl2}
\gamma^t(\tilde{\rho}, \rho) =
\int_{\Omega^t} d\rho(x) \int_{M \setminus \Omega^t} d\rho(y)\: 
\Big( f(x)\, \L\big(F(x),y \big) -  \L\big(x, F(y) \big)\, f(y) \Big)\:.
\eeq
In order to ensure that the nonlinear surface layer integral is well-defined and finite, we
need to make the following assumption.
\begin{Def} \label{defsla}
The measures~$\rho$ and~$\tilde{\rho}$ are {\bf{surface layer admissible}} if for all~$t \in \R$,
both integrals in~\eqref{osinl} are finite,
\[ \int_{\tilde{\Omega}^t} d\tilde{\rho}(x) \int_{M \setminus \Omega^t} d\rho(y)\: \L(x,y) < \infty\:,\qquad
\int_{\Omega^t} d\rho(x) \int_{\tilde{M} \setminus \tilde{\Omega}^t}  d\tilde{\rho}(y)\: \L(x,y) < \infty \:. \]
\end{Def}

We now recall the conservation law for the nonlinear surface layer integral.
A conservation law holds if the above surface layer integral vanishes when~$\Omega^t$ is replaced
by any compact set~$\Omega$. In order to analyze when this is the case, we
introduce a measure~$\nu$ on~$M$ and a measure~$\tilde{\nu}$ on~$\tilde{M}:=F(M)$ 
(the so-called {\em{correlation measures}}) by
\begin{align*}
d\nu(x) &:= \bigg( \int_{\tilde{M}} \L(x,y)\: d\tilde{\rho}(y) \bigg) \: d\rho(x) \\
d\tilde{\nu}(x) &:= \bigg( \int_M \L(x,y)\: d\rho(y) \bigg) \: d\tilde{\rho}(x) \:.
\end{align*}
Then~\eqref{osinl2} can be rewritten as
\begin{align*}
\gamma^t(\tilde{\rho}, \rho) &= \int_\Omega d\big(\Phi_*\tilde{\rho}\big)(x) \int_{M \setminus \Omega} \!\!\!\!
d\rho(y) \: \L\big(F(x),y \big) - \int_{\Omega} d\rho(x) \int_{M \setminus \Omega} \!\!\!\!d\big(\Phi_*\tilde{\rho} \big)(y)\:
\L\big(x,F(y)\big) \\
&= \int_\Omega d\big(\Phi_*\tilde{\rho}\big)(x) \int_{M}
d\rho(y) \: \L\big(F(x),y \big) - \int_{\Omega} d\rho(x) \int_{\tilde{M}} d\tilde{\rho}(y)\:
\L(x,y) \\
&= \int_\Omega d\big(\Phi_*\tilde{\nu}\big)(x) - \int_{\Omega} d\nu(x)
= \big( \Phi_* \tilde{\nu} - \nu \big)(\Omega) \:.
\end{align*}
We thus obtain the following result:
\begin{Prp} The surface layer integral~\eqref{osinl2} vanishes for every compact~$\Omega \subset M$
if and only if
\beq \label{mutilmu}
\nu = \Phi_* \tilde{\nu} \:.
\eeq
\end{Prp} \noindent
Exactly as explained in~\cite[Appendix~A]{fockbosonic}, the existence of a diffeomorphism~$\Phi$
follows from a general result in~\cite{greene-shiohama}.

Having fixed the foliations by
\beq \label{NtilN}
N_t = \partial \Omega^t \qquad \text{and} \qquad \tilde{N}_t = F(N_t) \:,
\eeq
there remains the freedom to perform diffeomorphisms on
each leaf~$\tilde{N}_t$, i.e.\ to perform the transformation
\[ (t,\x) \in \tilde{N}_t \rightarrow (t, \phi_t(\x)) \in \tilde{N}_t \]
for a family of diffeomorphisms~$\phi_t$ on~$N_t$, which for simplicity we assume to be smooth
in both~$t$ and~$\x$. By choosing these diffeomorphisms appropriately, one can change
the weight of the measure~$\rho$ arbitrarily (see~\cite{greene-shiohama}
and the constructions in~\cite[Appendix~A]{fockbosonic}).
In this way, we can arrange that
\beq \label{tilrho2}
\tilde{\rho} = F_* \rho \qquad \text{with} \qquad F \in C^\infty(M, \F \big) \:.
\eeq

\section{The Partition Function} \label{secZ}
As outlined in Section~\ref{secosinonlin}, the nonlinear surface layer integral~$\gamma^t(\tilde{\rho},\rho)$
was defined in the setting of causal variational principles by~\eqref{osinl}.
When working with causal fermion systems, however,
there is the complication that the vacuum system and the interacting system
are described by two causal fermion systems~$(\H, \F, \rho)$ and~$(\tilde{\H}, \tilde{\F}, \tilde{\rho})$ which
are defined on two different Hilbert spaces~$\H$ and~$\tilde{\H}$.
Therefore, in order to make sense of the nonlinear surface layer integral, we need to identify the
Hilbert spaces~$\H$ and~$\tilde{\H}$ by a unitary transformation denoted by~$V$,
\beq \label{Vident}
V : \H \rightarrow \tilde{\H} \qquad \text{unitary} \:.
\eeq
Then operators in~$\tilde{F}$ can be identified with operators in~$\F$ by the unitary transformation,
\beq \label{Ftrans}
\F = V^{-1}\, \tilde{\F}\, V \:.
\eeq
An important point to keep in mind is that this identification is not canonical, but it leaves the freedom
to transform the operator~$V$ according to
\beq \label{Vtrans}
V \rightarrow V \scrU \qquad \text{with} \qquad \scrU \in \Lin(\H) \text{ unitary} \:.
\eeq
For ease in notation, in what follows we always identify~$\H$ and~$\tilde{\H}$ via~$V$,
making it possible to always work in the Hilbert space~$\H$. Then the non-uniqueness of the identification
still shows up in the unitary transformation of the vacuum measure
\beq \label{rhotrans}
\rho \mapsto \scrU \rho \:,
\eeq
where~$\scrU \rho$ is defined by
\beq \label{Urhodef}
(\scrU \rho)(\Omega) := \rho \big( \scrU^{-1} \,\Omega\, \scrU \big) 
\qquad \text{for} \qquad \Omega \subset \F \:.
\eeq

Our strategy for dealing with this arbitrariness is to ``symmetrize'' the nonlinear surface layer integral
by integrating over the unitary transformations. 
In order to make the construction mathematically well-defined,
we only integrate over a compact subgroup~$\G$ of the (possibly infinite-dimensional)
unitary group~$\U(\H)$. In the applications, this subgroup must
be chosen sufficiently large. Moreover, in view of the physical applications in mind,
it seems sensible to assume that the unitary transformations are of relevance only
for the low-energy states, i.e.\ the
wave functions describing particles and anti-particles. With this in mind, we restrict attention to
the freedom to transform~$\tilde{\rho}$
with elements of a Lie subgroup~$\G$ of the unitary group on~$\H^\fermi$,
\beq \label{GsubU}
\G \subset \U(\H^\fermi)
\eeq
(where~$\H^\fermi$ is again the finite-dimensional subspace of~$\H$ introduced in Section~\ref{secELCFS}).
This freedom must be taken into account systematically in the subsequent constructions.
We remark that, in what follows, we may always choose~$\G=\U(\H^\fermi)$.
The only reason why in~\eqref{GsubU} we allow for~$\G$ to be a proper subgroup of the unitary group
is that in this way, the setting harmonizes with that used in the construction of the entropy 
for causal fermion systems in~\cite{entropy}.

When considering the nonlinear surface layer integral~\eqref{osinl}, we can treat the freedom~\eqref{rhotrans}
by integrating over~$\scrU \in \G$. To this end, we again assume that the measure~$\tilde{\rho}$
can be written in the form~\eqref{tilrho} with~$f \in C^\infty\big(M, \R^+ \big)$ and~$F \in C^\infty\big(M, \F \big)$.
In analogy to~\eqref{osinl2}, we define
\beq \label{osinlU}
\begin{split}
&\gamma^t(\tilde{\rho}, \scrU \rho) \\
&= \int_{\Omega^t} d\rho(x) \int_{M \setminus \Omega^t} d\rho(y)\: 
\Big( f(x)\, \L\big(F(x), \scrU y \scrU^{-1} \big) -  \L\big(\scrU x \scrU^{-1}, F(y) \big)\, f(y) \Big)\:.
\end{split}
\eeq
In order to ensure that the integrals are well-defined and finite,
we adapt Definition~\ref{defsla} as follows.

\begin{Def} \label{osiassume}
The measures~$\rho$ and~$\tilde{\rho}$ are {\bf{surface layer admissible}} if for all~$t \in \R$
and for all~$\scrU \in \G$, 
both terms in~\eqref{osinlU} are finite, i.e.\
\begin{align*}
\int_{\Omega^t} d\rho(x) \int_{M \setminus \Omega^t} d\rho(y)\: 
f(x)\, \L\big(F(x), \scrU y \scrU^{-1} \big) &< \infty \\
\int_{\Omega^t} d\rho(x) \int_{M \setminus \Omega^t} d\rho(y)\: 
\L\big(\scrU x \scrU^{-1}, F(y) \big)\, f(y) &< \infty \:.
\end{align*}
\end{Def}

\begin{Def}
The {\bf{symmetrized nonlinear surface layer integral}} is defined by
\[ \gamma^t_\G\big( \tilde{\rho}, \rho \big) = \fint_\G \gamma^t \big(\tilde{\rho}, \scrU \rho \big) \: d\mu_\G(\scrU) \:, \]
where~$d\mu_\G$ denotes the normalized Haar measure on~$\G$,
and~$\scrU \rho$ is the unitarily transformed measure~\eqref{Urhodef}.
\end{Def} \noindent
Writing the symmetrized nonlinear surface layer integral as
\[ \gamma^t_\G\big( \tilde{\rho}, \rho \big) = \gamma^t \big(\tilde{\rho}, \rho_\text{symm} \big) \:, \]
where~$\rho_\text{symm}$ is the symmetrized measure defined in analogy to~\eqref{Urhodef} by
\[ \rho_\text{symm}(\Omega) = \fint_\G \rho\big( \scrU^{-1} \,\Omega\, \scrU \big) \: d\mu_\G(\scrU) \:, \]
we can again apply Proposition~\ref{mutilmu} with~$\nu$ replaced by
its symmetrization~$\nu_\text{symm}$.
The existence of a diffeomorphism~$\Phi$ again follows from the general result in~\cite{greene-shiohama}
(for details see again~\cite[Appendix~A]{fockbosonic}).
In what follows, we always identify~$M$ and~$\tilde{M}$ via this diffeomorphism. Then~\eqref{mutilmu}
can be written as
\[ 
\bigg( \int_{\tilde{M}} \L\big(\scrU x \scrU^{-1} ,y)\: d\tilde{\rho}(y) \bigg) \: d\rho(x) =
\bigg( \fint_\G d\mu_\G(\scrU) \int_M d\rho(y)\: \L \big( x, \scrU y \scrU^{-1} \big)   \bigg) \: d\tilde{\rho}(x) \:. \]

As noted in~\cite{fockbosonic}, the Fock norm is to be identified with the exponential of~$\gamma^t(\rho, \tilde{\rho})$. In order to extend this connection to the fermionic Fock space,
it is important to integrate over~$\G$ {\em{after}} exponentiating. This motivates the following definition.

\begin{Def} \label{defZ}
The {\bf{partition function}} is defined by
\[ Z^t\big( \beta, \tilde{\rho} \big) = \fint_\G \exp \Big( \beta \,\gamma^t \big(\tilde{\rho}, \scrU \rho \big) \Big) \: d\mu_\G(\scrU) \]
with the unitarily transformed measure~$\scrU \rho$ as in~\eqref{Urhodef}.
\end{Def} \noindent
We point out that~$Z^t$ will in general not be conserved.
Moreover, $Z^t$ as well as the nonlinear surface layer integral
depend crucially on the choice of the foliations~\eqref{NtilN}.
We take the view that in the vacuum spacetime~$M$, the foliation~$(N_t)_{t \in \R}$ is given canonically
by the time function of an observer in an inertial frame in Minkowski space.
The choice of the foliation~$(\tilde{N}_t)_{t \in \R}$ of the interacting spacetime, however, is less
obvious. As worked out in~\cite{entropy}, this foliation can be obtained by a variational principle
in which~$Z^t$ is minimized under variations of~$\tilde{N}_t$ keeping~$\gamma^t$ fixed.
This minimization process is also shown in~\cite{entropy} to give a connection to a notion of entropy
and clarifies the significance of the parameter~$\beta$ and of the 
freedom in choosing the group~$\G$.

\section{Interacting Quantum Fields in a Surface Layer} \label{secqf}
The goal of this section is to describe the interacting measure~$\tilde{\rho}$ at time~$t$
by an interacting state~$\omega^t$ acting on the algebra of fields in the vacuum spacetime.
In preparation, we first introduce the field operators in the vacuum.
In order to relate the objects of the vacuum to the interacting spacetime, we
consider the partition function (see Definition~\ref{defZ}) with suitable insertions
(see Definition~\ref{defstate}).

\subsection{Field Operators in the Vacuum} \label{secalgebra}
We now introduce {\em{creation}} and {\em{annihilation operators}} operators in the spacetime~$M$ describing the vacuum (the connection to the algebra of observables and M{\o}ller operators will be explained in Section~\ref{secmoller}).
For the bosonic fields, this involves
the choice of a complex structure. In the setting of causal fermion systems, a canonical
complex structure~$J$ is induced on~$\Glin_{\rho, \sc}$ by the surface layer integrals~$(.,.)^t_\rho$
and~$\sigma^t_\rho$ (see~\eqref{sprod} and~\eqref{sigma} for~$\Omega=\Omega^t$;
for ease in notation we shall use the time parameter as upper index).
The construction, which is carried out in detail in~\cite[Section~6.3]{fockbosonic}, is summarized
as follows. We assume that the surface layer integral~$(.,.)^t_\rho$ defines a
scalar product on~$\Glin_{\rho, \sc}$. Dividing out the null space and forming the completion,
we obtain a real Hilbert space denoted by~$(\h^\R, (.,.)^t_\rho)$.
For the construction of~$J$, one assumes that~$\sigma^t_\rho$ is bounded relative to the scalar product~$(.,.)^t_\rho$. Then we can represent~$\sigma^t_\rho$ as
\[ 
\sigma^t_\rho(u, v) = (u,\, \mathscr{T}\, v)^t_\rho \:, \]
where~$\mathscr{T}$ is a uniquely determined bounded operator on the Hilbert space~$\h^\R$.
Assuming that~$\mathscr{T}$ is invertible, we set
\[ 
J := -(-\mathscr{T}^2)^{-\frac{1}{2}}\: \mathscr{T} \:. \]
Next, we complexify the Hilbert space~$\h^\R$ and denote
its complexification by~$\h^\C$. On this complexification, the operator~$J$ has eigenvalues~$i$
and~$-i$. The corresponding eigenspaces are referred to as the {\em{holomorphic}} and
{\em{anti-holomorphic subspaces}}, respectively.
We write the decomposition into holomorphic and anti-holomorphic components as
\[ 
\bv = \bv^\hol + \bv^\ah \:. \]
We also complexify the symplectic form to a {\em{sesquilinear}}
form on~$\h^\C$ (i.e.\ anti-linear in its first and linear in its second argument).
On the holomorphic jets we introduce a scalar product~$(.|.)^t_\rho$ by
\[ (.|.)^t_\rho := \sigma^t_\rho( \,.\,, J \,.\, ) \::\: \Gamma^\hol_\rho \times \Gamma^\hol_\rho \rightarrow \C \:. \]
Taking the completion gives a Hilbert space, which we denote by~$(\h, (.|.)^t_\rho)$.
This scalar product has the useful property that
\beq \label{imsigma}
\im ( u| v)^t_\rho = \im \sigma^t_\rho(u, J v) = \re \sigma^t_\rho(u, v) \:.
\eeq

The {\em{bosonic creation and annihilation operators}} are denoted by
\[ a^\dagger(z) \quad \text{and} \quad a(\overline{z}) \qquad \text{with} \qquad
z \in \h \]
(the overline in~$a(\overline{z})$ simply clarifies that this operator is anti-linear).
They satisfy the canonical commutation relations
\beq \label{CCR}
\big[ a(\overline{z}), a^\dagger(z') \big] = ( z | z' )^t_\rho \:,
\eeq
and all other operators commute,
\[ \big[ a(\overline{z}), a(\overline{z'}) \big] =  0 = \big[ a^\dagger(z), a^\dagger(z') \big]\:. \]

The {\em{fermionic creation and annihilation operators}} are denoted by
\[ \Psi^\dagger(\phi) \quad \text{and} \quad \Psi(\overline{\phi}) \qquad \text{for} \qquad
\phi \in \H^\fermi_{\rho, \sc} \:. \]
They satisfy the canonical anti-commutation relations
\beq \label{CAR}
\big\{ \Psi(\overline{\phi}), \Psi^\dagger(\phi') \big\} = \la  \phi | \phi' \ra^t_\rho \:,
\eeq
and all other operators anti-commute,
\[ \big\{ \Psi(\overline{\phi}), \Psi(\overline{\phi'}) \big\} = 0 = \big\{ \Psi^\dagger(\phi), \Psi^\dagger(\phi') \big\} \:. \]

\subsection{Bosonic Variations of the Nonlinear Surface Layer Integral} \label{secvarynonlin}
Our next construction step is to associate the above creation and annihilation operators to
surface layer integrals. Using these surface layer integrals as insertions in the partition function,
we shall obtain the $n$-point functions defining our quantum state (see Section~\ref{secstate}).
We begin with the bosonic insertions (the fermionic insertions will be constructed in the next
Section~\ref{secosifermi}).
In this case, we simply use the complexified jets~$z \in \h$ of the vacuum spacetime
(as introduced in Section~\ref{secalgebra}) in order to vary the measure~$\rho$.
As the insertions we take the corresponding first variations of the
surface layer integral~$\gamma^t(\tilde{\rho}, \scrU \rho)$ for fixed~$\scrU$,
\beq \label{Dzgamma}
D_{z} \gamma^t(\tilde{\rho}, \scrU \rho) \qquad \text{and} \qquad 
D_{\overline{z}} \gamma^t(\tilde{\rho}, \scrU \rho) \:.
\eeq
For multi-particle measurements, we take products of such variations, i.e.\ expressions of the form
\[ D_{z'_1} \gamma^t(\tilde{\rho}, \scrU \rho) \cdots D_{z'_p} \gamma^t(\tilde{\rho}, \scrU \rho)\;
D_{\overline{z}_1} \gamma^t(\tilde{\rho}, \scrU \rho) \cdots D_{\overline{z}_q} \gamma^t(\tilde{\rho}, \scrU \rho) \:. \]
 One can think of these variations as describing measurements by fictitious
observers who have chosen the identification of the Hilbert spaces~$\H$ and~$\tilde{\H}$ corresponding to~$\scrU$.

\subsection{Construction of the Fermionic Insertions} \label{secosifermi}
For the fermionic insertions, given the unitary transformation~$\scrU$ at time~$t$, we want to perform
a ``measurement'' in~$\H^{\fermi,t}_\rho$ which gives us the probability that a state described by a physical wave function~$\tilde{\psi}$ in the spacetime with interaction is occupied.
This makes it necessary to relate the wave functions in the interacting spacetime to
those in the vacuum. To this end, we again want to consider variations of the nonlinear surface layer integral.
In preparation, we need to introduce fermionic jet spaces. To this end,
we can make use of the fact that jets can be described by first variations of the
wave evaluation operator. Indeed, varying the formula
\[ F(x) = -\Psi(x)^* \Psi(x) \:, \]
a vector field~$\bv \in T_x \F$ can be written as
\beq \label{jetvdef}
\bv(x) = -\delta \Psi(x)^*\, \Psi(x) - \Psi(x)^*\, \delta \Psi(x) \:.
\eeq
A wave function~$\psi \in \H^{\fermi,t}_\rho$ by itself does not give rise to a variation of~$\Psi$.
As additional input we need a vector~$u \in \H^\fermi$, making it possible to vary the wave evaluation operator by
\[ \delta \Psi(x)\, v := \psi(x)\: \la u | v \ra_\H \:. \]
The corresponding jet~\eqref{jetvdef} takes the form
\[ 
\bv(x) = -|u\ra_\H\: \Sl \psi(x) \,|\, \Psi(x) \Sr_x - \Sl \Psi(x)^* \,|\, \psi(x) \Sr_x\: \la u | \:. \]
Taking linear combinations of these jets gives the vector space of {\em{fermionic jets}}
denoted by~$\Gamma^\fermi_\rho$.
On the fermionic jets one has a natural complex structure inherited from the complex structure
of~$\H^{\fermi}_\rho$ given by
\[ J\, \bv(x) = i\, |u\ra_\H\: \Sl \psi(x) \,|\, \Psi(x) \Sr_x
-i\, \Sl \Psi(x)^* \,|\, \psi(x) \Sr_x\: \la u | \:. \]
As a consequence, the corresponding holomorphic and anti-holomorphic components are given by
\beq \begin{split} \label{holaholfermi}
\bv^\hol(x) &= - \Sl \Psi(x)^* \,|\, \psi(x) \Sr_x\: \la u | \\
\bv^\ah(x) &= -|u\ra_\H\: \Sl \psi(x) \,|\, \Psi(x) \Sr_x \:.
\end{split}
\eeq
We shall always work with this complex structure. For ease in notation, we denote these jets by
\beq \label{jetfermi}
\bv^\hol = -\psi \, \la u | \quad \text{and} \quad \bv^\ah = -|u \ra\: \overline{\psi}
\qquad \text{with} \qquad \psi \in \H^{\fermi,t}_\rho,\; u \in \H^\fermi \:.
\eeq
Similar to~\eqref{Dzgamma}, we can now consider first variations of the nonlinear surface
layer integral,
\beq \label{Dfermigamma}
D_{\psi \, \la u |} \gamma^t(\tilde{\rho}, \scrU \rho) \qquad \text{and} \qquad 
D_{|u \ra\: \overline{\psi}} \gamma^t(\tilde{\rho}, \scrU \rho)
\eeq
(note that here we again vary the vacuum measure). In order to remove the
dependence on the vector~$u \in \H$, we choose an orthonormal basis~$(e_\ell)$ of~$\H^\fermi$
and sum over a bra/ket combination. This gives rise to the sesquilinear form
\beq \label{bbilinear}
b \::\: \H^{\fermi,t}_\rho \times \H^{\fermi,t}_\rho \rightarrow \C \:,\qquad
b(\psi, \phi) := \sum_\ell D_{|e_\ell \ra \, \overline{\psi}} \,\gamma^t(\tilde{\rho}, \scrU \rho)\;
D_{\phi \, \la e_\ell|} \,\gamma^t(\tilde{\rho}, \scrU \rho) \:.
\eeq
Obviously, this sesquilinear form is positive semi-definite and has finite rank bounded by~$f^\fermi$.
Therefore, we can represent it with relative to the scalar product~$\la .|. \ra^t_\rho$ on the extended Hilbert space,
\beq \label{Bdef}
b(\psi, \phi) = \la \psi \,|\, B\, \phi \ra^t_\rho \qquad \text{for all~$\psi, \phi \in \H^{\fermi,t}_\rho$}\:,
\eeq
where the operator~$B \in \Lin(\H^{\fermi,t}_\rho)$ is positive semi-definite and has rank at most~$f^\fermi$.
We denote the orthogonal projection onto its image by
\beq \label{piUdef}
\pi^{\fermi,t} \::\: \H^{\fermi,t}_\rho \rightarrow B(\H^{\fermi,t}_\rho) \:\subset\: \H^{\fermi,t}_\rho\:.
\eeq

We next consider a subspace~$I \subset \H^{\fermi, t}_\rho$ of dimension~$p$.
Diagonalizing the operator~$\pi^{\fermi,t}$ on this subspace, we obtain an orthonormal
basis~$\psi_1, \ldots, \psi_p$ of~$I$ with
\beq \label{fermiinsert}
\la \psi_i \,|\, \pi^{\fermi,t}\, \psi_j \ra^t_\rho = \nu_i\: \delta_{ij} \:.
\eeq
We can interpret the vectors~$\psi_1, \ldots, \psi_p$ as the wave functions detected in
simultaneous one-particle measurements with corresponding probabilities~$\nu_1, \ldots, \nu_p$.
Hence the product~$\nu_1 \cdots \nu_p$ can be interpreted as the probability of the $p$-particle measurement.
Clearly, this product of eigenvalues coincides with the determinant of~$\pi^{\fermi, t}$ on~$I$,
\[ \nu_1 \cdots \nu_p = \det \big(\pi_I \,\pi^{\fermi,t}|_I \big) \:, \]
where~$\pi_I : \H^{\fermi,t}_\rho \rightarrow I \subset \H^{\fermi,t}_\rho$ is the orthogonal projection.
This determinant can also be expressed directly in terms of the sesquilinear form~$\la \,.\, |\,\pi^{\fermi, t}\,. \,\ra^t_{\tilde{\rho}, \scrU}$
by working with totally anti-symmetrized wave functions,
\[ \nu_1 \cdots \nu_p = \frac{1}{p!} \sum_{\sigma, \sigma' \in S_p} (-1)^{\sign(\sigma)+\sign(\sigma')}
\la \psi_{\sigma(1)} \,|\, \pi^{\fermi, t}\, \psi_{\sigma'(1)} \ra^t_\rho
\cdots \la \psi_{\sigma(p)} \,|\, \pi^{\fermi,t}\, \psi_{\sigma'(p)} \ra^t_\rho \:, \]
where~$S_p$ is the symmetric group.

\subsection{The State Induced by the Causal Fermion System at Fixed Time} \label{secstate}
We denote the unital $*$-algebra generated by the field operators of Section~\ref{secalgebra}
by~$\A$. A {\em{state}}~$\omega$ is a linear mapping from the algebra
which is positive, i.e.\
\[ \omega : \A \rightarrow \C \qquad \text{with} \qquad
\omega(A^* A) \geq 0 \quad \text{for all~$A \in \A$}\:. \]
In this section we will prove the main result of this paper, which states that
the interacting measure~$\tilde{\rho}$ gives rise to a distinguished state at time~$t$
(see Definition~\ref{defstate} and Theorem~\ref{thmpositive}).

Before entering the constructions, we recall the main assumptions entering our result.
First, we assume that the vacuum spacetime is of Minkowski type (see Section~\ref{secminktype}).
Second, the linearized solutions in the vacuum should be well-behaved in the sense that
the surface layer inner product~$( .,. )^t_\rho$ is positive definite, that the
symplectic form is non-degenerate, and that these bilinear forms can be related to each other
by a bounded operator~$\T$ (see the beginning of Section~\ref{secalgebra}).
All these assumptions have been verified explicitly in examples (for details see~\cite{action}).

Next, we assume that the interacting measure can be written as the push-forward of the
vacuum measure~\eqref{tilrho}. 
A further simplifying assumption is to restrict attention to locally compact variations
(see Section~\ref{seccfscvp}). For the fermionic variations, this amounts to varying the physical
wave functions only in a finite-dimensional subspace~$\H^\fermi \subset \H$ (see~\eqref{Hfermidef}).
Likewise, for the identifications of the Hilbert spaces we work with a finite-dimensional
Lie subgroup~$\G$ of the unitary group (see~\eqref{GsubU}).
Finally, we assume that the nonlinear surface layer
integral which compares the measures~$\rho$ and~$\tilde{\rho}$ is finite for all~$\scrU \in \G$
(see Definition~\ref{osiassume}).
These assumptions are technical simplifications which should eventually be relaxed.
Doing so would be an important objective, making it necessary to use infinite-dimensional
analysis as partly explored in~\cite{banach}. Moreover, one would have to deal with the
non-compactness of the unitary group. Pursuing this program seems worthwhile but challenging.

We now enter our construction. Using linearity and the canonical commutation and anti-commutation relations~\eqref{CCR} and~\eqref{CAR}, the state is determined by defining how it acts on
a product of operators where all creation operators are on the left and all
annihilation operators are on the right. We define this expectation value by
inserting the variations of the nonlinear surface layer integral in Section~\ref{secvarynonlin}
as well as the fermionic expectation values constructed in Section~\ref{secosifermi}
into the integrand of the partition function (see Definition~\ref{defZ}).

For notational clarity, we refer to a linear functional~$\omega^t$ on the field algebra~$\A$
as a {\em{pre-state}}, i.e.\
\beq \label{prestate}
\omega^t \::\: \A \rightarrow \C \quad \text{complex linear} \:.
\eeq
The pre-state is a {\em{state}} if it is positive in the sense that
\beq \label{state}
\omega^t \big( A^* A \big) \geq 0 \qquad \text{for all~$t \in \R$ and~$A \in \A$}\:.
\eeq
We first define a pre-state and prove positivity afterward.
\begin{Def} \label{defstate}
The {\bf{pre-state}}~$\omega^t$ at time~$t$ is defined by
\begin{align*}
&\omega^t\Big( a^\dagger(z'_1) \cdots a^\dagger(z'_p)\:\Psi^\dagger(\phi'_1) \cdots \Psi^\dagger(\phi'_{r'}) \;
a(\overline{z_1}) \cdots a(\overline{z_q}) \: \Psi(\overline{\phi_1}) \cdots \Psi(\overline{\phi_r}) \Big) \\
&:= \frac{1}{Z^t\big( \beta, \tilde{\rho} \big)} \: \delta_{r' r}\:\frac{1}{r!} \sum_{\sigma, \sigma' \in S_{r}}
(-1)^{\sign(\sigma)+\sign(\sigma')} \\
&\quad\;\: \times
\int_\G e^{\beta\, \gamma^t(\tilde{\rho}, \scrU \rho)}\;
\la \phi_{\sigma(1)} \,|\, \pi^{\fermi, t}\, \phi'_{\sigma'(1)} \ra^t_\rho
\cdots \la \phi_{\sigma(r)} \,|\, \pi^{\fermi, t} \, \phi'_{\sigma'(r)} \ra^t_\rho  \\
&\qquad\quad \times 
D_{z'_1} \gamma^t(\tilde{\rho}, \scrU \rho) \cdots D_{z'_p} \gamma^t(\tilde{\rho}, \scrU \rho)\;
D_{\overline{z}_1} \gamma^t(\tilde{\rho}, \scrU \rho) \cdots D_{\overline{z}_q} \gamma^t(\tilde{\rho}, \scrU \rho) \; d\mu_\G(\scrU) \:.
\end{align*}
\end{Def} \noindent
We point out that this pre-state vanishes unless there are as many fermionic creation as annihilation operators.
This means that our pre-state is an eigenstate of the fermionic particle number operator
(where, as usual, we count anti-particles with a minus sign).

We next prove positivity.
\begin{Thm} \label{thmpositive} The pre-state~$\omega^t$ is a {\bf{state}}.
\end{Thm}
\Proof Our task is to prove the inequality~\eqref{state}.
A general element~$A \in \A$ of the field algebra
can be written as a finite linear combination of operator products of the form
\begin{align*}
A = \sum_{p,p',q,q'} \sum_k c(p,p',q,q', k)&
\: a^\dagger(z'_{\alpha,1}) \cdots a^\dagger(z'_{\alpha, p'})\: a(\overline{z_{\alpha,1}}) \cdots a(\overline{z_{\alpha,p}}) \\
&\!\!\!\!\!\!\!\!\times \: \Psi^\dagger(\phi'_{\alpha,1}) \cdots \Psi^\dagger(\phi'_{\alpha, q'})\: \Psi(\overline{\phi_{\alpha,1}}) \cdots \Psi(\overline{\phi_{\alpha,q}}) \:,
\end{align*}
where~$\alpha:=(p,p',q,q', k)$ is a multi-index. We form the product~$A^* A$ and multiply out.

Let us first treat the fermionic field operators. We get a nonzero contribution to the expectation value
only if there are as many creation as annihilation operators. Moreover, we obtain pairings between
a creation and an annihilation operator according to
\beq \label{oneminuspi}
\Psi^\dagger(\phi) \, \Psi(\overline{\psi}) \;\text{ gives }\; \la \psi \,|\, \pi^{\fermi, t}\, \phi \ra^t_\rho \quad \text{and} \quad
\Psi(\overline{\psi})\, \Psi^\dagger(\phi) \;\text{ gives }\; \la \psi \,|\, (\1-\pi^{\fermi, t})\, \phi \ra^t_\rho \:.
\eeq
There are pairings between operators within~$A$ and within~$A^*$.
For those pairings, we simply replace the operator products by the corresponding expectation values.
Then it remains to consider pairings in which one field operator in~$A$ is combined with one
field operator in~$A^*$.

Next, in order to treat the bosonic field operators, we use the commutation relations such as to bring all
the creation operators to the left and all the annihilation operators to the right.
This gives rise to pairings according to
\[ a(\overline{z})\: a^\dagger(z') \;\text{ gives }\; (z | z')^t_\rho \:, \]
where always one annihilation operator~$A^*$ is combined with one creation operator in~$A$.
After having performed all these commutations, we end up with a product of bosonic field
operators of the form as in Definition~\ref{defstate}, where each creation and annihilation operator
gives rise to an insertion~$D_{z} \gamma^t$ and~$D_{\overline{z}} \gamma^t$, respectively.

After these transformations, the expectation value can be written as follows,
\begin{align*}
&\omega^t(A^* A) = \int_\G d\mu_\G(\scrU) \sum_{p,q,r=0}^\infty \overline{T^{i_1 \ldots i_p,\: j_1 \cdots j_q}_{k_1 \cdots k_r}}(\scrU) \;
T^{l_1 \ldots l_p,\: m_1 \cdots m_q}_{n_1 \cdots n_r}(\scrU) \: (z_{k_1} | z_{n_1})^t_\rho \cdots (z_{k_r} | z_{n_r})^t_\rho \\
&\times \la \psi_{i_1} \,|\, \pi^{\fermi, t}\, \psi_{l_1} \ra^t_\rho \cdots \la \psi_{i_p} \,|\, \pi^{\fermi, t}\, \psi_{l_p} \ra^t_\rho\;
\la \psi_{j_1} \,|\, (\1-\pi^{\fermi, t})\, \psi_{m_1} \ra^t_\rho \cdots \la \psi_{j_q} \,|\, (\1-\pi^{\fermi, t})\, \psi_{m_q} \ra^t_\rho \:,
\end{align*}
where the functions~$T^{i_1 \ldots i_p,\: j_1 \cdots j_q}_{k_1 \cdots k_r}$ are symmetric in the lower ``bosonic'' indices
and totally anti-symmetric in the upper ``fermionic'' indices~$i_1, \ldots, i_p$ and~$j_1,\ldots, j_q$.
The integrand is non-negative because all the involved inner products are positive semi-definite.
This concludes the proof.
\QED

\subsection{Constructing Representations of the Field Algebra}
\subsubsection{The GNS Representation}
The state~$\omega^t$ can be represented abstractly using the GNS construction,
as we now briefly recall (for details see for example~\cite[Section~1.6]{arveson-Cstar}
or~\cite[Section~5.1.3]{khavkine-moretti}). On the unital $*$-algebra~$\A$ we introduce the
sesquilinear form
\[ \la A | A' \ra := \omega^t(A^* A')\::\: \A \times \A \rightarrow \C \:. \]
Using the positivity of the state~$\omega^t$ (see Theorem~\ref{thmpositive}), this sesquilinear
form is positive semi-definite. Dividing out the null space and forming the completion, we obtain
a Hilbert space~$(\Fock, \la .|. \ra)$. Since the multiplication in the algebra gives a representation
of~$\A$ on itself, we also obtain a canonical representation of~$\A$ on~$\Fock$.
Moreover, by definition,
the vector~$\Phi$ representing the identity in~$\A$ has the property that
\[ \la \Phi \,|\, A\, \Phi \ra = \omega^t(\1^*\, A\, \1) = \omega^t(A) \:. \]
Thus in the above representation, the state is realized as the expectation value of~$\Phi$.
This is the abstract GNS representation.

We point out that, unless the state is quasi-free, the GNS representation need {\em{not}} be
a Fock representation. In order to remedy the situation, we now proceed by constructing a Fock representation
on the free Fock space.

\subsubsection{Representation on the Free Fock Space}
We now explain how the state~$\omega^t$ can be represented on the Fock space
of the free field theory.

In preparation, we build up the Fock space for the vacuum measure~$\rho$.
In order to get the connection to the usual vacuum state used in Dirac theory,
we need to implement the fermionic frequency splitting and the Dirac sea picture.
To this end, we introduce the orthogonal projections
\[ \pi_- \::\: \H^{\fermi,t}_\rho \rightarrow \H^\fermi \subset \H^{\fermi,t}_\rho \qquad \text{and} \qquad
\pi_+ = (\1-\pi_i) \::\: \H^{\fermi,t}_\rho \rightarrow (\H^\fermi)^\perp \]
and set
\[ \Psi^\dagger_\pm(\psi) := \Psi^\dagger( \pi_\pm \psi) \qquad \text{and} \qquad
\Psi_\pm(\overline{\psi}) := \Psi( \overline{\pi_\pm \psi}) \:. \]
Obviously, these operators satisfy the anti-commutation relations
\beq \label{CARsplit}
\big\{ \Psi_s(\overline{\phi}), \Psi^\dagger_{s'}(\phi') \big\} = \la  \phi \,|\, \pi_{s,s'}\, \phi' \ra^t_\rho
\eeq
(and all other fermionic operators anti-commute). Moreover,
\[ \Psi^\dagger = \Psi^\dagger_+ + \Psi^\dagger_- \qquad \text{and} \qquad
\Psi = \Psi_+ + \Psi_- \:. \]
The subscript~$\pm$ can be understood as a generalization of the usual splitting of the solution space
into solutions of positive and negative frequencies, respectively.

We introduce the vacuum~$|0\ra \in \Fock$ as the vector with the property
\beq \label{vacdef}
0 = a(\overline{z})\,|0\ra = \Psi_+(\overline{\psi})\,|0\ra = \Psi^\dagger_-(\psi)\,|0\ra 
\qquad \text{for all~$z \in \h$ and~$\psi \in \H^{\fermi,t}_\rho$} \:.
\eeq
We now consider the vector space generated by~$A |0\ra$ with~$A \in \A$.
The scalar product between such vectors is determined by~\eqref{vacdef},
the commutation relations~\eqref{CCR} and the anti-commutation relations~\eqref{CARsplit}.
Taking the completion gives the Fock space~$(\Fock, \la .|. \ra_\Fock)$.

Before going on, we make a few explanatory remarks. In view of~\eqref{vacdef}, the
Fock space is built up by acting on the vacuum state by the operators
\beq \label{create}
a^\dagger(z) \:,\qquad \Psi^\dagger_+(\psi) \quad \text{and} \quad \Psi_-(\overline{\psi}) \:.
\eeq
These operators can be viewed as generating the Fock space by creating particles and anti-particles.
The annihilation operator~$\Psi_-$ of the vectors in~$\H^\fermi$ is to be interpreted as the
operator creating a hole in the Dirac sea. By renaming the creation and annihilation operators
according to~$\Psi_- \leftrightarrow \Psi^\dagger_-$, one gets the usual notation in
quantum field theory (in order to avoid confusion, we shall not adopt this convention here).
Since all the bosonic field operators commute with all the fermionic field operators, the Fock space is a tensor product
of the subspaces generated by the bosonic and fermionic operators,
\[ \Fock := \Fock^\bose \otimes \Fock^\fermi \:. \]
The vectors in the bosonic subspace~$\Fock^\bose$ can be identified with the
symmetric tensor product of~$\h$, i.e.\
\[ \Fock^\bose = \bigoplus_{n=0}^\infty \Fock^{\bose, n} \qquad \text{and} \qquad
\Fock^{\bose, n} := (\h^n)_\symm \:, \]
where the index~$\symm$ stands for the total symmetrization, i.e.\
\[ \big( \psi_1 \otimes \cdots \otimes \psi_n \big)_\symm := \frac{1}{n!} \sum_{\sigma \in S_n}
\psi_{\sigma(1)} \otimes \cdots \otimes \psi_{\sigma(n)} \:. \]
More details on this way of constructing the bosonic Fock space can be found in~\cite[Section~7.1]{fockbosonic}.
The vectors of the fermionic subspace~$\Fock^\fermi$, on the other hand, are products of totally antisymmetric
wave functions describing the particles and the anti-particles, i.e.\ vectors of the form
\[ \big(\psi_1 \wedge \cdots \wedge \psi_p\big) \wedge \big( \phi_1 \wedge \cdots \wedge \phi_q\big) \]
with~$\psi_i \in (\H^\fermi)^\perp$ and~$\phi_i \in \H^\fermi$,
endowed with the scalar product induced by~$\la .|. \ra^t_\rho$.

Our task is to construct an operator~$\sigma^t$ on~$\Fock$ with the property that
\beq \label{fockrep}
\omega^t(A) = \tr_\Fock(\sigma^t A) \qquad \text{for all~$A \in \A$} \:.
\eeq
Bringing all the operators in~\eqref{create} to the left with the commutation and anti-commu\-ta\-tion relations,
it clearly suffices to satisfy the condition~\eqref{fockrep} for all operators~$A$ of the form
\beq \label{Areorder}
\begin{split}
A &= a^\dagger(z'_1) \cdots a^\dagger(z'_{r'}) 
\;\Psi^\dagger_+(\psi'_{+,1}) \cdots \Psi^\dagger_+(\psi'_{+,{p_+}}) \;
\Psi_-(\overline{\psi_{-,1}}) \cdots \Psi_-(\overline{\psi_{-,{p_-}}}) \\
&\quad\:\times a(\overline{z_1}) \cdots a(\overline{z_r}) \;\Psi_+(\overline{\psi_{+,1}}) \cdots \Psi_+(\overline{\psi_{+,{q_+}}})\;
\Psi^\dagger_-(\psi'_{-,1}) \cdots \Psi^\dagger_-(\psi'_{-,{q_-}}) \:.
\end{split}
\eeq
Regarding~$A$ as describing a measurement, the parameters~$r,r', p_\pm$ and~$q_\pm$
tell us how many particles and anti-particles are involved in the process.
In order to clarify issues of convergence, we first consider the case that the process involves
only a finite number of particles and anti-particles.
\begin{Def} The state~$\omega^t$ involves a {\bf{finite number of particles and anti-particles}}
if there is~$N \in \N$ such that~$\omega^t(A)=0$ for all~$A$ of the form~\eqref{Areorder} whenever
one of the parameters~$r,r',p_\pm$ or~$q_\pm$ exceeds~$N$.
\end{Def}

\begin{Thm} \label{thmrep}
Assume that the states~$\omega^t$ involves a finite number of particles or anti-particles. Then there is
a density operator~$\sigma^t$ on~$\Fock$ such that~\eqref{fockrep} holds.
\end{Thm}
\Proof
We first explain the idea for a purely bosonic state. The fermions will be treated below with similar methods.
Thus, disregarding the fermions, the operator~$A$ in~\eqref{Areorder} simplifies to
\beq \label{Abos}
A = a^\dagger(z'_1) \cdots a^\dagger(z'_{r'}) \; a(\overline{z_1}) \cdots a(\overline{z_r}) \:.
\eeq
The assumption that~$\sigma^t$ involves a finite number of particles and anti-particles means that for
some~$R>r$,
\[ \omega^t(A) = 0 \qquad \text{if~$r>R$}\:. \]
We refer to~$r$ as the number of particles involved in the measurement and to~$R$ as the maximal
number of such particles.
Our goal is to show that there is a linear operator~$\sigma^t$ which involves at most~$R$ particles and
represents~$\omega^t$ for measurements involving~$R$ particles, i.e.
\beq \label{Rrel}
\omega^t(A) - \tr_\Fock(\sigma^t A) = 0 \qquad \text{if~$r>R-1$}\:.
\eeq
Once this has been achieved, we can proceed inductively for decreasing values of~$R$ until~$R=0$.
Adding the operators~$\sigma^t$ of the finite number of induction steps gives the desired operator~$\sigma^t$
satisfying~\eqref{fockrep} for all~$A$ of the form~\eqref{Abos}.

In order to construct an operator~$\sigma^t$ which satisfies~\eqref{Rrel}, we first choose an orthonormal
basis~$(z_i)$ of~$\h$. Then the operators
\[ A^{j_1, \ldots, j_R}_{i_1, \ldots, i_{r'}} := a^\dagger(z_{i_1}) \cdots a^\dagger(z_{i_{r'}}) \; a(\overline{z_{j_1}}) \cdots a(\overline{z_{j_R}}) \]
with indices~$i_1 \leq i_2 \leq \cdots \leq i_{r'}$ and~$j_1 \leq j_2 \leq \cdots \leq j_R$ form a basis
of the operators~\eqref{Abos} with~$R=r$. Setting
\begin{align*} 
\sigma^t &= \sum_{r'} \omega^t \big( A^{j_1, \ldots, j_R}_{i_1, \ldots, i_{r'}} \big)
\bigg( \prod_{\ell'=1}^{r'} \frac{1}{(m(i_{\ell'})!)^\frac{1}{m(i_{\ell'})}} \bigg)
\bigg( \prod_{\ell=1}^{R} \frac{1}{(m(j_\ell)!)^\frac{1}{m(j_\ell)}} \bigg) \\
&\qquad\qquad\qquad\qquad \times |a^\dagger(z_{j_1}) \cdots a^\dagger(z_{j_R}) | 0 \ra
\la 0 | a(\overline{z_{i_1}}) \cdots a(\overline{z_{i_{r'}}}) | \:,
\end{align*}
where~$m(i_\ell)$ and~$m(j_\ell)$ denote the multiplicities of the corresponding indices,
a direct computation using the canonical commutation relations shows that~\eqref{Rrel} holds
(note that the powers~$1/m_{i_\ell}$ and~$1/m_{j_\ell}$ are needed in order to compensate for the
fact that the corresponding factors appear several times). This concludes the proof for a purely bosonic state.

The general case with fermions can be proved similarly by proceeding inductively for decreasing values of~$r$
as well as of the fermionic particle numbers~$q_+$ and~$q_-$. Moreover, the operator~$\sigma^t$
must be complemented by fermionic operators of the form
\begin{align*}
&| \Psi^\dagger_+(\psi_{+,i_1}) \cdots \Psi^\dagger_+(\psi_{+,i_{q_+}})\;
\Psi_-(\overline{\psi_{-,j_1}}) \cdots \Psi_-(\overline{\psi_{-,j_{q_-}}}) \,|\, 0 \ra \\
&\quad\:\times  \la 0 \,|\, \Psi_+(\overline{\psi_{+,k_1}}) \cdots \Psi_+(\overline{\psi_{+,k_{p_+}}}) \;
\Psi^\dagger_-(\psi_{-,l_1}) \cdots \Psi^\dagger_-(\psi_{-,l_{p_-}}) | \:,
\end{align*}
where the fermionic indices are strictly increasing due to the nilpotence of the fermionic creation
and annihilation operators. Consequently, there are no multiplicities for the fermionic states.
After these obvious modifications, the above arguments again go through.
\QED

In order to extend this result to a general state~$\omega^t$, one can decompose the state
into a series
\[ \omega^t = \sum_{N=1}^\infty \omega_N^t \:, \]
where each~$\omega^t_N$ involves at most~$N$ particles or anti-particles
(the~$\omega^t_N$ could be constructed for example by projection to subspaces of the Fock space).
With the help of Theorem~\ref{thmrep},
we can represent each~$\omega^t_N$ by a linear operator~$\sigma^t_N$. Taking their sum,
\beq \label{sigmasum}
\sigma^t := \sum_{N=1}^\infty \sigma^t_N \:,
\eeq
one obtains the desired representation of~$\omega^t$ by a density operator~$\sigma^t$.
However, one should keep in mind that the sum in~\eqref{sigmasum} can be understood only formally.
We cannot expect that this sum converges for any state~$\omega^t$.
This corresponds to the well-known problem in quantum field theory of inequivalent Fock representations
(see for example~\cite{klaus+scharf1, nenciu+scharf} for this problem
in the context of quantum fields in a classical external field).
Here we shall not enter the analysis of such convergence issues.
Nevertheless, the representation constructed in Theorem~\ref{thmrep} seems useful at least
in a perturbative treatment.

\subsection{Realizing the Insertions as Functional Derivatives} \label{secderivative}
It is a natural question whether, in analogy to the path integral formulation of quantum field theory,
the above insertions in the state~$\omega^t$ (see Definition~\ref{defstate})
can be realized as variational derivatives of the partition function, i.e.\ symbolically
\[ \omega^t(\cdots) = \frac{1}{\beta^k\, Z^t\big( \beta, \tilde{\rho} \big)} \underbrace{D \cdots D}_{\text{$k$ derivatives}}
Z^t\big( \beta, \tilde{\rho} \big) \:. \]
The short answer is yes, up to rather subtle technical issues.
We now explain the connection in more detail.

The bosonic insertions are variational derivatives of the exponent of the partition function.
Therefore, in order to obtain the desired product of insertions, one can take corresponding
variational derivatives of~$Z^t$, but one must make sure that each derivative acts on the exponential. 
This could be arranged for example by restricting attention to the
highest order in~$\beta$, i.e.\ symbolically
\beq \label{highestorder}
\omega^t(\cdots) = \frac{1}{\beta^k\, Z^t\big( \beta, \tilde{\rho} \big)} \underbrace{D \cdots D}_{\text{$k$ derivatives}}
Z^t\big( \beta, \tilde{\rho} \big)\; \Big( 1 + \O \big( \beta^{-1} \big) \Big) \:.
\eeq
The disadvantage of this method is that one would have to control the error term by
showing that the higher derivatives of the nonlinear surface layer integral are negligible.
This analysis has not yet been carried out. This is why we prefer to define the bosonic
state with insertions, each of which involves one functional derivative.

We next consider the fermionic insertions.
Exactly as explained for the bosonic insertions before~\eqref{highestorder},
the first variations of the nonlinear surface layer integral~\eqref{Dfermigamma}
can again be realized by corresponding variational derivatives of~$Z^t$, up to higher orders in~$\beta^{-1}$.
But again, the resulting error term has not yet been analyzed.
Carrying out the sum over~$\ell$, also the bilinear form in~\eqref{bbilinear} could be
realized as a variational derivatives of~$Z^t$.
But there is the subtle issue that, instead of working with this bilinear form,
the fermionic insertions were constructed from the expectation values of the
projection operator~$\pi^{\fermi, t}$ defined in~\eqref{piUdef}.
Working with this projection operator has the advantage that also the
operator~$(\1 - \pi^{\fermi, t})$ is positive semi-definite. This was made use of
in the proof of the positivity of the state (see~\eqref{oneminuspi}).
If we knew that the operator~$B$ defined by~\eqref{bbilinear} and~\eqref{Bdef}
has norm at most one, then we could replace the operator~$\pi^{\fermi, t}$
in~\eqref{fermiinsert} by~$B$, making it possible to express the fermionic insertions
directly in terms of variational derivatives of the nonlinear surface layer integral.
In order to establish this bound on the norm of~$B$, one would need the inequality
\[ \sum_\ell D_{|e_\ell \ra \, \overline{\psi}} \,\gamma^t(\tilde{\rho}, \scrU \rho)\;
D_{\psi \, \la e_\ell|} \,\gamma^t(\tilde{\rho}, \scrU \rho)
\leq \la \psi | \psi \ra^t_\rho \qquad \text{for all~$\psi \in \H^{\fermi, t}_\rho$}\:. \]
Although we expect that this inequality should hold in many applications, it has not yet been proven.
For this reason, we need to work with the projection
operator~$\pi^{\fermi, t}$.

\subsection{The Algebra of Observables and the Quantum M{\o}ller Map} \label{secmoller}
We finally put our construction into the context of algebraic quantum field theory
by discussing the algebras of observables and making the connection to M{\o}ller operators.

For clarity, we first introduce the {\em{algebra of observables of the vacuum}},
denoted by~$\scrA_\rho$. For the bosonic operators, as in~\cite[Section~7.2]{fockbosonic}
to every~$\u \in \J^*_0$ we associate a corresponding symmetric operator~$\hat{\Phi}(\u)$.
The bosonic algebra of observables is the free algebra generated by these operators
satisfying the commutation relations
\beq \label{CCRobserve}
\big[ \hat{\Phi}(\u), \hat{\Phi}(\v) \big] 
= i \: \la \u, G \,\v \ra_{L^2(M)}\: \1 \:,
\eeq
where~$G$ is the bosonic Green's operator~\eqref{Jlinscdef}.
Making use of~\eqref{sigmaform} and~\eqref{imsigma}, the bosonic observables can be expressed in terms of the creation and annihilation operators by (for details see~\cite[Section~7.2]{fockbosonic})
\[ 
\hat{\Phi}(\u) = a(\overline{G \u}) + a^\dagger( G \u ) \:. \]
Likewise, to every~$\psi \in \scrW_0$ we associate the operators~$\hat{\Psi}^*(\psi)$
and~$\hat{\Psi}(\overline{\psi})$. The fermionic algebra of observables is the algebra
generated by these operators which satisfies the anti-commutation relations
\beq \label{CARobserve}
\{ \hat{\Psi}(\overline{\phi}), \hat{\Psi}^*(\psi) \big\}
= \bra \phi \,|\, k \,\psi \ket\: \1 \:,
\eeq
where~$k$ is the fermionic Green's operator~\eqref{kdef} and~$\bra .|. \ket$ is the Krein inner product~\eqref{krein}.
All other fermionic operators anti-commute,
\[ \big\{ \hat{\Psi}(\overline{\phi}), \hat{\Psi}(\overline{\phi'}) \big\} = 0 = \big\{
\hat{\Psi}^*(\phi), \hat{\Psi}^*(\phi') \big\} \:. \]
The algebra of observables~$\scrA_\rho$ is defined as the unital $*$-algebra generated freely by the
bosonic and fermionic algebras
(which means that all bosonic field operators commute with all fermionic field operators).
The fermionic observables are expressed in terms of the creation and
annihilation operators simply by
\[ \hat{\Psi}^*(\psi) = \Psi^\dagger(k \psi) \qquad \text{and} \qquad 
\hat{\Psi}(\overline{\psi}) = \Psi(\overline{k \psi}) \:. \]
Indeed, making use of~\eqref{kkrel}, for any~$\phi, \psi \in \scrW_0$,
\begin{align*}
\big\{ \Psi(\overline{k \phi}), \Psi^\dagger(k \psi) \big\}
= \la k \phi \,|\, k \psi \ra^t_\rho =  \bra \phi \,|\, k\, \psi \ket \:,
\end{align*}
giving agreement with~\eqref{CARobserve}.
The algebra~$\scrA_\rho$ corresponds precisely to the algebra of observables
for quasi-free quantum field theories in algebraic quantum field theory.

We finally point out that the causal fundamental solutions~$G$ and~$k$
entering the canonical (anti-)commutation relations~\eqref{CCRobserve} and~\eqref{CARobserve}
are causal in the sense that they vanishes if their two arguments are spacelike separated
and sufficiently far apart
(for technical details see~\cite[Section~5.3 and Section~6]{linhyp} and~\cite[Section~6.5]{dirac}).
In this way, the causal structure of spacetime is incorporated in the algebra~$\scrA_\rho$.

\section{Outlook} \label{secoutlook}
\subsection{The Algebra of Interacting Fields}
In order to get further connection to the algebraic formulation, it would be desirable to
also have an algebra of observables~$\scrA_{\tilde{\rho}}$ for the interacting spacetime.
The commutation and anti-commutation relations would involve the interacting Green's operators.
At present, it is not clear how the algebra~$\scrA_{\tilde{\rho}}$ could be constructed.
Part of the problem is that the interacting spacetime may involve observables which do not
correspond to the degrees of freedom described by the field operators in~$M$.
A more detailed discussion of this point can be found in~\cite{mix}.
In view of these difficulties, here we are more modest and identify the operators in~$\scrA_{\tilde{\rho}}$
with operators in~$\scrA_\rho$ only at fixed time~$t$. More precisely, in view of the time slice axiom
(or, more computationally, by taking the Hamiltonian time evolution in the Heisenberg picture),
the algebras are uniquely determined by all field operators in a time strip~$[t_0, t_0+\Delta t]$,
where~$\Delta t>0$ can be chosen arbitrarily small. Therefore, identifying the operators of~$\scrA_{\tilde{\rho}}$
and~$\scrA_\rho$ in this time strip, we obtain an algebra homomorphism
\beq \label{iota}
\iota^t \::\: \scrA_{\tilde{\rho}} \rightarrow \scrA_\rho \:.
\eeq
Identifying the algebras only at a fixed time has the advantage that the equal time commutation
and anti-commutation relations of the vacuum can also be used for the interacting operators.
In other words, we describe the interacting quantum fields at time~$t$ with the field operators
of the vacuum. The state of the interacting system is described at time~$t$ by the
state~$\omega^t$.

A homomorphism~\eqref{iota} identifying the interacting algebra with the algebra in the vacuum
appears in algebraic quantum field theory as the so-called {\em{quantum M{\o}ller map}}. 
For details and the broader context we refer to the recent paper~\cite{drago-hack-pinamonti}
and the references therein.

\subsection{The Dynamics of the Quantum Fields} \label{secqdyn}
In the previous section, at a given time~$t$ we constructed a state~$\omega^t$
describing the interaction as encoded in the measure~$\tilde{\rho}$.
By performing this construction at different times, one can recover the dynamics of the system.
However, it is a-priori not clear whether this dynamics can be described by a time evolution
operator acting on the state
\[ {\mathfrak{L}}^t_{t_0} \::\: \omega^{t_0} \rightarrow \omega^t \:. \]
Going one step further, one would like to realize this time evolution by a unitary transformation~$U^t_{t_0}$
on the Fock space, i.e.
\[ U^t_{t_0} : \Fock \rightarrow \Fock \text{ unitary} \qquad \text{and} \qquad
\sigma^t = U^t_{t_0}\, \sigma^{t_0}\, (U^t_{t_0})^{-1} \:. \]
The question whether such time evolution operators exist and how they looks like will be the
objective of the forthcoming paper~\cite{mix}.
%
%

\subsection{A Few Refinements of the Construction} \label{secrefined}
In this section we mention a few modifications and refinements of the above constructions
which may be useful in the applications.

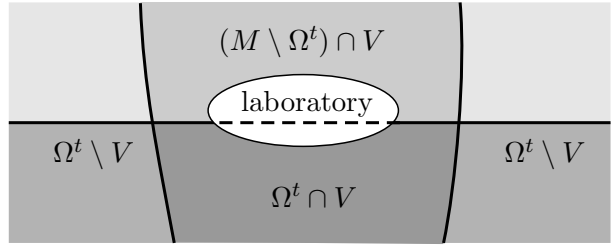
\begin{figure}[tb]
\psscalebox{1.0 1.0} 
{
\begin{pspicture}(0,25.994967)(8.02,29.235035)
\definecolor{colour0}{rgb}{0.7019608,0.7019608,0.7019608}
\definecolor{colour1}{rgb}{0.9019608,0.9019608,0.9019608}
\definecolor{colour2}{rgb}{0.6,0.6,0.6}
\definecolor{colour3}{rgb}{0.8,0.8,0.8}
\psframe[linecolor=colour0, linewidth=0.04, fillstyle=solid,fillcolor=colour0, dimen=outer](8.01,27.620033)(0.01,26.020035)
\psframe[linecolor=colour1, linewidth=0.04, fillstyle=solid,fillcolor=colour1, dimen=outer](8.01,29.220034)(0.01,27.620033)
\pspolygon[linecolor=colour2, linewidth=0.02, fillstyle=solid,fillcolor=colour2](1.93,27.615034)(6.0,27.595034)(5.94,27.175034)(5.92,26.835033)(5.85,26.365034)(5.78,26.015034)(2.2,26.045034)(2.04,26.965034)
\pspolygon[linecolor=colour3, linewidth=0.02, fillstyle=solid,fillcolor=colour3](1.77,29.215034)(5.99,29.215034)(6.02,28.905035)(6.03,28.525034)(6.01,28.065035)(5.99,27.635035)(1.93,27.635035)(1.83,28.385035)(1.79,28.715034)
\psline[linecolor=black, linewidth=0.04](0.01,27.620033)(8.01,27.620033)
\psbezier[linecolor=black, linewidth=0.04](1.76,29.210033)(1.8,28.100035)(1.98,27.250034)(2.21,26.0200341796875)
\psbezier[linecolor=black, linewidth=0.04](6.01,29.220034)(6.09,28.290033)(5.95,26.780035)(5.79,26.0200341796875)
\psellipse[linecolor=black, linewidth=0.02, fillstyle=solid, dimen=outer](3.925,27.782534)(1.275,0.4875)
\psline[linecolor=black, linewidth=0.04, linestyle=dashed, dash=0.17638889cm 0.10583334cm](2.81,27.620033)(5.21,27.620033)
\rput[bl](0.6,27){$\Omega^t \setminus V$}
\rput[bl](6.6,27){$\Omega^t \setminus V$}
\rput[bl](3.5,26.5){$\Omega^t \cap V$}
\rput[bl](2.8,28.5){$(M \setminus \Omega^t) \cap V$}
\rput[bl](3.1,27.7){laboratory}
\end{pspicture}
}
\caption{Localizing the state.}
\label{figlocstate}
\end{figure}

The quantum state defined in Section~\ref{secstate} was defined globally in space.
For most applications, however, it seems more suitable to define a state describing only a {\em{bounded
spatial region}}, which can be thought of as our laboratory or the subsystem of our present universe
accessible to measurements. In order to implement this picture, we choose a subset~$V \subset M$
of spacetime such that its intersection with the surface layer contains the spatial region of interest (see
Figure~\ref{figlocstate}).
We define the {\em{localized nonlinear surface layer integral}}~$\gamma^{t,V}$ by
restricting all the integrals in the nonlinear surface layer integral~\eqref{osinlU} to~$V$,
\beq \label{osinlUV}
\begin{split}
\gamma^t_V(\tilde{\rho}, \scrU \rho) :=
\int_{\Omega^t \cap V} &d\rho(x) \int_{(M \setminus \Omega^t) \cap V} d\rho(y) \\
&\times \Big( f(x)\, \L\big(F(x), \scrU y \scrU^{-1} \big) -  \L\big(\scrU x \scrU^{-1}, F(y) \big)\, f(y) \Big)\:.
\end{split}
\eeq
We next define the {\em{localized state}}~$\omega^t_V$ by replacing all surface layer integrals
in Definition~\ref{defstate} (also those in the insertions) by corresponding localized surface layer integrals.
Working with the localized state is particular useful for analyzing the asymptotics when the
total spatial volume tends to infinity (corresponding to the limit~$f \rightarrow \infty$),
while the spatial region described by~$V$ is kept fixed. This analysis will be carried out in
detail in~\cite{mix}.

Another refinement concerns the choice of the group~$\G$ which we integrate over
in the partition function and the state (see Definitions~\ref{defZ} and~\ref{defstate}).
On the present abstract level of the present paper,
we did not need to be specific and simply chose~$\G$ as a subgroup of~$\U(\H^\fermi)$
(see~\eqref{GsubU}). However, when working out the state more concretely
(as will be done in~\cite{mix}), one would like to consider the limiting case
when~$\H^\fermi$ exhausts~$\H$ and~$\G$ goes over to the whole unitary group~$\U(\H)$.
In this limiting case, one must take into account the fact that the unitary group
includes the undesirable unitary transformations describing symmetries of Minkowski space,
like time translations and spatial translations.
One method of removing these undesirable unitary transformations from the group integral,
which is explored in~\cite{entropy}, is to impose additional side conditions.
Alternatively, we now propose to remove the freedom of performing translations in Minkowski
space by inserting into the integrands of the partition function and the state the factor~$e^{\alpha \T^t_V}(\tilde{\rho}, \scrU \rho)$ with
\beq
\begin{split}
\T^t_V(\tilde{\rho}, \scrU \rho) &:= \bigg( 
\int_{\Omega^t \cap V} \int_{\Omega^t \cap V}  + \int_{(M \setminus \Omega^t) \cap V}
\int_{(M \setminus \Omega^t) \cap V} \bigg) \\
&\qquad\qquad\qquad \times
f(x)\, \big| F(x)\, \scrU y \scrU^{-1} \big|^2\: d\rho(x)\: d\rho(y)
\end{split}
\eeq
(where the absolute value is again the spectral weight~\eqref{sw}). Thus the
{\em{localized partition function}}~$Z_V$ and the {\em{localized state}}~$\omega^t_V$ are defined
similar as in Definitions~\ref{defZ} and~\ref{defstate} by
\begin{align*}
 Z^t_V\big( \alpha, \beta, \tilde{\rho} \big) &:= \fint_\G e^{\alpha \T^t_V(\tilde{\rho}, \scrU \rho) +  \beta \,\gamma^t_V (\tilde{\rho}, \scrU \rho)} \: d\mu_\G(\scrU) \\
\omega^t_V \big( \cdots \big)
&:= \frac{1}{Z^t_V\big( \alpha, \beta, \tilde{\rho} \big)}
\fint_\G e^{\alpha \T^t_V(\tilde{\rho}, \scrU \rho) +  \beta \,\gamma^t_V (\tilde{\rho}, \scrU \rho)} \: 
\big( \cdots \big) \:d\mu_\G(\scrU) \:,
\end{align*}
where we abbreviated the insertions by~$(\cdots)$. Choosing~$\alpha$
sufficiently large (and~$V$ as a set of finite measure), the main contribution to the group integral is obtained when
the functional~$\T^t_V$ is maximal.
For unitary transformations describing translations in Minkowski space, the maximum is
attained when the sets~$\Omega^t \cap V$ and~$(M \setminus \Omega^t) \cap V$
coincide with the corresponding transformed sets. In this way, the translation freedom is fixed.

We conclude this section by explaining a method which allows to extract additional information on the interacting
measure~$\tilde{\rho}$ by building in separate phases for the bra and ket components of the insertions.
To this end, we shall also modify the partition function as introduced in Definition~\ref{defZ} by working
with a double integral over two
unitary transformations~$\scrU_<$ and~$\scrU_>$ which act on bra and ket, respectively.
More precisely, given two causal fermion systems~$(\H, \F, \rho)$ and~$(\tilde{\H}, \tilde{\F}, \tilde{\rho})$,
we again identify the Hilbert spaces~$\H$ and~$\tilde{\H}$~\eqref{Vident}, leaving us with the
freedom~\eqref{Vtrans}. We consider~$\F$ as a subset of the set~$\F^\C$
all operators of rank at most~$2n$,
\[ \F \subset \F^\C := \big\{ A \in \Lin(\H) \:\big|\: \text{$A$ has rank at most $2n$} \big\}\:. \]
We extend~$\rho$ and~$\tilde{\rho}$ by zero to~$\F^\C$.
The causal Lagrangian is readily extended to the~$\F^\C \times \F^\C$. Indeed, for any~$x, y \in \F^\C$,
the operator product~$xy$ is again an operator of rank at most~$2n$.
Denoting its non-trivial eigenvalues again by~$\lambda^{xy}_1, \ldots \lambda^{xy}_{2n}$,
we can work again with the definition~\eqref{Lagrange}
(note that, due the absolute values, the Lagrangian is again real-valued and non-negative).
Consequently, also the nonlinear surface layer integral~\eqref{osinl}
may be defined for measures~$\rho$ and~$\tilde{\rho}$ on~$\F^\C$. Next, we generalize~\eqref{rhotrans} and~\eqref{Urhodef} to the transformation
\[ \rho \mapsto T_{\scrU_<, \scrU_>} \rho \:, \]
where
\[ \big( T_{\scrU_<, \scrU_>} \rho \big)(\Omega) := \rho \big( \scrU_>^{-1} \,\Omega\, U_< \big) \]
(the subscripts of the operators~$\scrU_>$ and~$\scrU_<$ remind that these operators
will act on the bra and ket component of the insertions, respectively).
Clearly, the measure~$T_{\scrU_<, \scrU_>} \rho$ may be supported also on non-symmetric operators, i.e.
\[ \supp T_{\scrU_<, \scrU_>} \rho \subset \F^\C \:. \]
The corresponding nonlinear surface layer is introduced in analogy to~\eqref{osinlU} by
\beq \label{osinlUref}
\begin{split}
\gamma^t(\tilde{\rho}, T_{\scrU_<, \scrU_>} \rho) = &\int_{\Omega^t} d\rho(x) \int_{M \setminus \Omega^t} d\rho(y) \\
&\;\;\times
\Big( f(x)\, \L\big(F(x), \scrU_> \,y\, \scrU_<^{-1} \big) -  \L\big(\scrU_> \,x\, \scrU_<^{-1}, F(y) \big)\, f(y) \Big)\:.
\end{split}
\eeq
Moreover, we define the {\em{refined partition function}} in analogy to Definition~\ref{defZ} by
\[ Z^t_{\text{\rm{ref}}}\big( \beta, \tilde{\rho} \big) = \fint_\G d\mu_\G \big(\scrU_< \big)  \fint_\G d\mu_\G \big( \scrU_> \big)
\exp \Big( \beta \,\gamma^t \big(\tilde{\rho}, T_{\scrU_<, \scrU_>}  \rho \big) \Big) \:. \]

Next, we want to define the refined pre-state in analogy to Definition~\ref{defstate} by placing
fermionic and bosonic insertions into the integrand. However, we want to distinguish
between variations in the bra and ket components.
In preparation for the bosonic insertions, we recall that a jet~$\bv$ (again without scalar component)
can be described by a 
corresponding variation of the wave evaluation operator,
\[ \bv(x) = -\big( \delta_\bv \Psi(x)^* \big) \Psi(x) - \Psi(x)^* \big( \delta_\bv \Psi(x) \big) \:. \]
We remark that the variation~$\delta_\bv \Psi$ is unique only up to local gauge transformations.
Here this freedom is not of relevance; it could be removed for example using the gauge-fixing procedure in~\cite{gaugefix}.
Similarly, the holomorphic and anti-holomorphic components, denoted by~$z$ and~$\overline{z}$,
can be written as
\begin{align*}
z(x) &= -\Big( \delta_z \big(\Psi(x)^* \big) \Big) \Psi(x) - \Psi(x)^* \big( \delta_z \Psi(x) \big) \\
\overline{z}(x) &= -\Big( \delta_{\overline{z}} \big(\Psi(x)^* \big) \Big) \Psi(x) - \Psi(x)^* \big( \delta_{\overline{z}} \Psi(x) \big)
\end{align*}
(note that the derivatives of both bra and ket are complex linear in~$z$ and~$\overline{z}$, respectively).
We now modify these relations by introducing variational derivatives which act either on bra or on ket.
More precisely, we set
\begin{align*}
z_<(x) &:= - \Psi(x)^* \big( \delta_z \Psi(x) \big) \\
\overline{z}_>(x) &:= - \Big( \delta_{\overline{z}} \big( \Psi(x)^*\big) \Big) \Psi(x) \:.
\end{align*}
The corresponding jet derivatives are also denoted by
\[ D_z^< = D_{z_<} \qquad \text{and} \qquad D_{\overline{z}}^> = D_{\overline{z}_<} \:. \]

For the fermionic insertions, we modify the construction in Section~\ref{secosifermi}.
In the variational derivatives of the surface layer integrals
in~\eqref{bbilinear}, we build in the transformations~$U_<$ or~$U_>$ by defining
\beq \label{bdefnew}
b(\psi, \phi) := \sum_\ell D^>_{| U_>^{-1} U_< e_\ell \ra \, \overline{\psi_i}} \,\gamma^t \big(\tilde{\rho},  T_{\scrU_<, \scrU_>}  \rho \big)\;
D^<_{\psi_j \, \la e_\ell|} \,\gamma^t \big(\tilde{\rho}, T_{\scrU_<, \scrU_>}  \rho \big)
\eeq
(where we used again the notation~\eqref{jetfermi}; the factor~$U_>^{-1} U_<$ avoids a relative unitary
transformation of the fermionic bra and ket vectors).
%
%
The resulting operator~$B$ defined by~\eqref{Bdef}
will in general not be symmetric. For this reason, it seems preferable not to introduce~$\pi^{\fermi, t}$
as a projection operator, but to work directly with the variational derivatives of the nonlinear surface layer
integrals by setting
\beq \label{pidefmodify}
\pi^{\fermi,t} \::\: \H^{\fermi,t}_\rho \rightarrow \H^{\fermi,t}_\rho \:,\qquad
\la \psi \,|\, \pi^{\fermi,t} \,\phi \ra^t_\rho := b(\psi, \phi)
\eeq
with the bilinear form~$b$ as given by~\eqref{bdefnew}.

\begin{Def} \label{defstateref}
The {\bf{refined pre-state}}~$\omega^t_{\text{\rm{ref}}}$ at time~$t$ is defined by
\begin{align*}
&\omega^t_{\text{\rm{ref}}}\Big( a^\dagger(z'_1) \cdots a^\dagger(z'_p)\:\Psi^\dagger(\phi'_1) \cdots \Psi^\dagger(\phi'_{r'}) \;
a(\overline{z_1}) \cdots a(\overline{z_q}) \: \Psi(\overline{\phi_1}) \cdots \Psi(\overline{\phi_r}) \Big) \\
&:= \frac{1}{Z^t_{\text{\rm{ref}}}\big( \beta, \tilde{\rho} \big)} \: \delta_{r' r}\:\frac{1}{r!} \sum_{\sigma, \sigma' \in S_{r}}
(-1)^{\sign(\sigma)+\sign(\sigma')} \\
&\quad\: \times
\fint_\G d\mu_\G \big(\scrU_< \big)  \fint_\G d\mu_\G \big( \scrU_> \big)\: e^{\beta\, \gamma^t(\tilde{\rho},T_{U_<, U_>} \rho)}\;
\la \phi_{\sigma(1)} \,|\, \pi^{\fermi, t}\, \phi'_{\sigma'(1)} \ra^t_\rho
\cdots \la \phi_{\sigma(r)} \,|\, \pi^{\fermi, t} \, \phi'_{\sigma'(r)} \ra^t_\rho  \\
&\qquad\quad \times 
D^<_{z'_1} \gamma^t(\tilde{\rho}, T_{U_<, U_>} \rho) \cdots D^<_{z'_p} \gamma^t(\tilde{\rho}, T_{U_<, U_>} \rho)\;
D^>_{\overline{z}_1} \gamma^t(\tilde{\rho}, T_{U_<, U_>} \rho) \cdots D^>_{\overline{z}_q} \gamma^t(\tilde{\rho}, \scrU \rho) \:.
\end{align*}
\end{Def}

We point out that the refined pre-state is not necessarily positive. The reason is that the
insertions~$D^<_{z} \gamma^t(\tilde{\rho}, T_{U_<, U_>} \rho)$
and~$D^>_{\overline{z}} \gamma^t(\tilde{\rho}, T_{U_<, U_>} \rho)$ are not
complex conjugates of each other, because taking the complex conjugate also amounts to
interchanging the two unitary transformations~$U_<$ and~$U_>$,
\[ \overline{D^<_{z} \gamma^t(\tilde{\rho}, T_{U_<, U_>} \rho)}
= D^>_{\overline{z}} \gamma^t(\tilde{\rho}, T_{U_>, U_<} \rho) \:. \]
Similarly, taking the complex conjugate of the fermionic insertion involves
flipping~$U_<$ and~$U_>$.
As a consequence, the positivity of the refined pre-state needs to be verified in the applications.
The detailed construction is postponed again to the follow-up paper~\cite{mix}.

\Thanks{{{\em{Acknowledgments:}} We would like to thank Claudio Dappiaggi, J\"urg Fr\"ohlich,
Marco Oppio, Claudio Paganini, Moritz Reintjes and J\"urgen Tolksdorf for helpful discussions.
We are grateful to the referees for valuable feedback and suggestions.
We would like to thank the ``Universit\"atsstiftung Hans Vielberth'' for support.
N.K.'s research was also supported by the NSERC grant RGPIN~105490-2018.

\providecommand{\bysame}{\leavevmode\hbox to3em{\hrulefill}\thinspace}
\providecommand{\MR}{\relax\ifhmode\unskip\space\fi MR }
\providecommand{\MRhref}[2]{%
  \href{http://www.ams.org/mathscinet-getitem?mr=#1}{#2}
}
\providecommand{\href}[2]{#2}

\end{document}